\documentclass[pdflatex,sn-mathphys-num,iicol]{sn-jnl}

\usepackage{graphicx}%
\usepackage{multirow}%
\usepackage{amsmath,amssymb,amsfonts}%
\usepackage{amsthm}%
\usepackage{mathrsfs}%
\usepackage[title]{appendix}%
\usepackage{xcolor}%
\usepackage{textcomp}%
\usepackage{manyfoot}%
\usepackage{booktabs}%
\usepackage{algorithm}%
\usepackage{algorithmicx}%
\usepackage{algpseudocode}%
\usepackage{listings}%

\usepackage{soul}
\usepackage{xr}
\usepackage{lipsum}
\usepackage[table]{xcolor}
\usepackage{caption}

\newcommand{\ud}{\text{d}}

\renewcommand{\phi}{\varphi}

\renewcommand{\rm}[1]{\mathrm{#1}}
\newcommand{\mbf}[1]{\mathbf{#1}}

\newcommand{\lrp}[1]{\left( #1 \right)} 
\newcommand{\lrb}[1]{\left[ #1 \right]} 
\newcommand{\lrc}[1]{\left\{ #1 \right\}} 

\newcommand{\partDer}[2]{\frac{\partial #1}{\partial #2}}
\newcommand{\partDerH}[2]{{\partial #1}/{\partial #2}}

\newcommand{\rev}[1]{\textcolor{black}{#1}}

\theoremstyle{thmstyleone}%
\theoremstyle{thmstyletw`o}%

\theoremstyle{thmstylethree}%

\begin{document}
	
	\title[]
	{Kinetics of Droplet Cloaking and Wetting Ridge Growth on Lubricated Polymer Brushes}
	

	\author[1]{Antonio Torregrosa Abell\'an}
	
	\author[2,3]{Enqing Liu}
	
	\author[3]{Vincent Siekman}
	
	\author[3]{Frieder Mugele}
	
	\author*[1]{Friederike Schmid}\email{friederike.schmid@uni-mainz.de}
	
	\author*[1]{Rodrique G. M. Badr}\email{rodrique.badr@physik.uni-freiburg.de}

	\affil[1]{\orgdiv{Institut f\"ur Physik}, \orgname{Johannes Gutenberg-Universit\"at Mainz}, \orgaddress{\street{Staudingerweg 7-9}, \postcode{55099}, \city{Mainz}, \country{Germany}}}
	
	\affil[2]{\orgdiv{College of Integrated Circuits \& Micro-Nano Electronics}, \orgname{Fudan University}, \orgaddress{\street{Handan Road 220}, \postcode{200433}, \city{Shanghai}, \country{China}}}
	
	\affil[3]{\orgdiv{Physics of Complex Fluids, MESA+ Institute}, \orgname{University of Twente}, \orgaddress{\street{PO box 217}, \postcode{7500AE}, \city{Enschede}, \country{The Netherlands}}}
	
	
	
	
	
	

	
	\abstract{We investigate the kinetics of wetting ridge growth and droplet cloaking on lubricant-infused polymer brushes using a combination of experiments, molecular dynamics simulations, and theoretical modeling. We focus on three representative systems: DMSO-water on hexadecane-swollen PLMA (D-H), water on hexadecane-swollen PLMA (W-H), and water on PDMS (W-S). The dynamics are governed by the interplay between interfacial thermodynamics, brush elasticity, and transport of lubricant within the brush. Ridge growth is accompanied by the formation of depletion zones both beneath and outside the drop. This leads to a progressive slowdown governed by the need to transport lubricant through the brush. At sufficiently high swelling, we observe local separation of oil from the brush within the ridge, providing an additional mechanism for lubricant depletion. To rationalize these observations, we develop a continuum diffusion model based on the free energy of the brush and its coupling to the contact line. The model quantitatively captures the growth of the wetting ridge at intermediate and late times, demonstrating that the kinetics are largely controlled by diffusive transport within the brush.}

	
	
	
	\maketitle
	
	\section{Introduction}
	
	Polymer brushes are widely used to modify solid surfaces,
	with applications ranging from adhesion reduction and 
	lubrication enhancement
	\cite{wright2016oil,daniel2019hydration,teisala2020grafting,li2022vapor,chen2023omniphobic,liu2025tuning}
	to diagnostics \cite{heggestad2020pursuit,
		li2021responsive} and biosensing \cite{wang2024pyroelectric}, among
	others. In many of these contexts, the interaction of polymer brushes with
	liquids, particularly liquid drops, is essential. Polymer brushes 
	exhibit several advantageous properties. \rev{They can adapt to the presence of drops \cite{butt2018adaptive,hartmann2024drops} or respond to external stimuli \cite{mukherji2020smart, schubotz2023influence} and change their wetting properties accordingly. In addition, drops on polymer brushes can have} low contact angle
	hysteresis, while the coatings themselves are highly durable, and retain lubricants very well
	\cite{lhermerout2019contact,teisala2020grafting}. The last property is
	central for applications such as ball bearings and self cleaning
	surfaces, where the brush acts as a reservoir for oil lubricants
	that reduce friction \cite{teisala2020grafting,
		badr2024dynamics,buonaiuto2026polymer}. The desirable properties of
	polymer brushes interacting with liquid drops arise from the
	interplay between brush elasticity, interfacial properties, 
	and the presence of liquid lubricant within the brush.
	
	Much effort has been devoted to understanding the interaction of
	liquid drops with lubricant-infused surfaces
	\cite{lafuma2003superhydrophobic,rykaczewski2013mechanism,wexler2015shear,daniel2017oleoplaning,daniel2020quantifying,wong2020adaptive,keiser2017drop}
	and elastomers \cite{style2013universal,zhao2018growth,cai2021fluid}.
	More recently, particular attention has been given to
	lubricant-infused polymer brushes \cite{teisala2020grafting, badr2022cloaking,
		buonaiuto2025thermally}. A key question in such systems
	is whether the lubricant is extracted from the brush in the presence of a drop. 
	Interfacial forces at the three-phase contact line can drive oil
	separation, leading to the formation of a two-phase wetting
	ridge \cite{cai2021fluid} or even complete cloaking of the drop
	\cite{kreder2018film,naga2021water,badr2022cloaking}. These processes
	contribute to lubricant depletion on the substrate and, consequently,
	to the degradation of surface functionality
	\cite{rykaczewski2013mechanism,wexler2015shear,teisala2020grafting}.
	Moreover, the presence of a liquid phase in the wetting ridge can
	influence drop dynamics by modifying dissipation mechanisms
	\cite{hauer2023phase}. 
	
	In this work, we investigate the kinetics of lubricant redistribution
	during the growth of wetting ridges and the formation of cloaking
	layers.  We adopt a combined experimental, theoretical, and
	simulation-based approach. Experimentally, we study aqueous Dimethyl
	sulfoxide (DMSO) drops deposited on Poly-lauryl methacrylate (PLMA)
	brushes swollen with hexadecane and track the time evolution of the
	wetting ridge. These investigations are complemented by molecular
	dynamics simulations of an analogous system.  We further use simulations
	to investigate oil separation in water-PLMA brush systems and to quantify
	the progression of the nanoscale cloaking layer in a water-PDMS system. Finally, motivated by the
	observed redistribution of lubricant, we develop a theoretical
	framework based on a diffusion equation derived from the free energy
	of the brush and the contact line.

	
	\section{Model and Methods}
	\label{sec:sim}
	
	We consider systems in which a liquid drop is deposited on a
	lubricated polymer brush. The primary experimental systems consist of
	PLMA brushes swollen with hexadecane and PDMS brushes swollen with
	silicone oil. In both cases, the \rev{monomers of} lubricant and grafted chains can be
	regarded as chemically identical, and we treat them as such in our
	simulations. \rev{ 
		However, in 
		both experiments and simulation, the number of monomers per oil chain $N_o$ is much smaller than that of grafted chains $N_B$. In particular, we have $N_B > N_o^2$ for all of our systems, which implies that the lubricant acts as an athermal solvent for the grafted chains \cite{rubinstein2003polymer}.} While water is the primary liquid of interest, aqueous
	DMSO solutions (DMSO40 corresponding to 40 wt \% concentration by
	weight) on PLMA are also considered to probe lower surface tension
	conditions.  Table \ref{tab:experimentalSurfTen} shows the
	experimental values for the relevant interfacial tensions (values for
	PDMS extracted from Ref.  \citenum{badr2022cloaking}).
	
	\begin{table}
		\centering
		\begin{tabular}{ |c|c|c|c| } 
			\hline
			(mN/m) & air & PDMS & Hexadecane \\
			\hline
			air &  \cellcolor{black} & 21 & $28.76 \pm 0.18$ \\
			\hline
			water &  $72.6 \pm 0.8$ & 40 & $50.4 \pm 0.4$ \\
			\hline
			DMSO40 & $62.3 \pm 0.9$ & \cellcolor{black} & $31.4 \pm 0.6$ \\
			\hline
		\end{tabular}
		\caption{Experimental interfacial tension values of the different interfaces that appear in our systems of interest.}
		\label{tab:experimentalSurfTen}
	\end{table}
	
	Simulation parameters are chosen to reproduce the ratios of
	interfacial tensions in the experimental systems, ensuring realistic
	contact angles and spreading behavior.

	
	\subsection{Simulations} \label{sec:simulationModel}
	
	We employ a coarse-grained model system comprising a polymer brush,
	free oil chains, and a droplet of liquid particles coexisting with its
	vapor. Brush polymers and oil
	molecules are modeled as chains of identical beads, denoted $p$,
	connected by springs with the spring potential
	$U_{\text{bond}}=\rev{\frac{1}{2}}k(r_{ij}-r_0)^2$. 
	Liquid molecules are modeled as single
	isolated beads (type $l$).  To model non-bonded interactions and the
	coupling to a heat bath at a given temperature $T$, we use the
	Many-body Dissipative Particle Dynamics (MDPD) coarse-grained model
	and thermostat. The DPD thermostat has the advantage that the
	dissipative and random forces are pairwise interactions and momentum
	conserving allowing for hydrodynamic phenomena
	\citep{hoogerbrugge1992simulating,espanol1995hydrodynamics,marsh1997static},
	while the multi-body force element allows for modeling the coexistence
	of two phases \citep{pagonabarraga2002mdpd, trofimov2002mdpd,
		warren2003vapor}.  The forces take the following form:
	
	\begin{align}
		&\mbf{F}_{ij}=\mbf{F}_{ij}^C+\mbf{F}_{ij}^D+\mbf{F}_{ij}^R\\
		&\mbf{F}_{ij}^C=\left(A_{ij}w^C(r_{ij})+B(\bar{\rho}_i+\bar{\rho}_j)\tilde{w}^C(r_{ij})\right)\hat{\mbf{r}}_{ij} 
		\rev{ + \mbf{F}_{ij}^{\text{bond}}} \\
		&\mbf{F}_{ij}^D=-\zeta w^C(r_{ij})^2(\hat{\mbf{r}}_{ij}.\mathbf{v}_{ij})\hat{\mbf{r}}_{ij}\\
		&\mbf{F}_{ij}^R=\sqrt{2\zeta k_B T}w^C(r_{ij})\theta_{ij}\hat{\mbf{r}}_{ij}\\
		&w^C(r_{ij})=\begin{cases}
			\bigg(1-\frac{r_{ij}}{r_c}\bigg)~~&r_{ij}\leq r_c\\
			0~~&r_{ij}>r_c
		\end{cases} \\
		&\tilde{w}^C(r_{ij})=\begin{cases}
			\bigg(1-\frac{r_{ij}}{r_d}\bigg)~~&r_{ij}\leq r_d\\
			0~~&r_{ij}>r_d
		\end{cases}\\
		&\bar{\rho}_i=\sum_{j\neq i}\frac{15}{2\pi r_d^3}\tilde{w}^C(r_{ij})^2
	\end{align}
	
	In the above equations, $\mbf{F}_{ij}^C$ is the conservative force
	contribution where $A_{ij}<0$ is the strength of the \rev{nonbonded}
	attractive part,
	$B>0$ is the strength of the density dependent repulsion\rev{, and
		the spring forces
		$\mbf{F}_{ij}^{\text{bond}} =-  k (r_{ij}-r_0) \hat{\mbf{r}}_{ij}$
		act only between neighbor monomers on a chain}. $B$ must
	have the same value for all pairs of particles for the forces to be
	conservative as shown by the no-go theorem of MDPD
	\citep{warren2013no}. We also have
	$\mathbf{r}_{ij}=\mathbf{r}_i-\mathbf{r}_j$, and
	$\hat{\mbf{r}}_{ij}=\mathbf{r}_{ij}/r_{ij}$. $\mbf{F}_{ij}^D$ and $\mbf{F}_{ij}^R$ are
	the dissipative and random force contributions respectively, where
	$\zeta$ is the drag coefficient,
	$\mathbf{v}_{ij}=\mathbf{v}_i-\mathbf{v}_j$, $k_B$ and $T$ are
	Boltzmann’s constant and the temperature respectively, and
	$\theta_{ij}$ is an uncorrelated Gaussian distributed random variable
	with zero mean and unit variance. $w^C$ and $\tilde{w}^C$ are weight
	functions, $\tilde{\rho}_i$ is a weighted density, and finally $r_c$
	and $r_d$ are cutoff radii which set the range of the forces. The
	reason for introducing two cutoff radii is that the range of the
	density-dependent repulsion must be smaller than that of the
	attraction, $r_d<r_c$, (with $A_{ij} < 0$ and $B > 0$) in order to
	obtain liquid-vapor coexistence \citep{warren2003vapor}.
	
	The polymer brush consists of end-grafted chains of length $N_B$. The chains are grafted to a purely
	repulsive surface modeled using the Weeks-Chandler-Anderson (WCA)
	potential \citep{weeks1971role}. The oil is modeled as free chains of
	length $N_o$. As already mentioned, monomers in the free oil chains
	and the grafted brush chains are all taken to be of the same species
	($p$) and therefore have the same interaction parameters among each
	other and with the liquid particles.
	
	The simulations are performed \rev{at constant number of particles, volume, and temperature, i.e.} in the \rev{canonical} NVT ensemble, using periodic
	boundary conditions in all directions. The unit of energy is set by
	$k_BT=1$, the unit of length by the cutoff distance of the DPD
	attraction $r_c=1$, and the mass unit by $m=1$ for all species. The
	unit of time can then be defined as $\tau=\sqrt{\frac{m
			r_c^2}{k_BT}}$. In the following, all quantities are given in these
	units. The model parameters were chosen to be the same as in our
	previous work \cite{badr2022cloaking}, with the DPD parameters
	$r_d=0.8;~\zeta=4.5;~ B=40$ (see below for a discussion of the choices
	of the parameters $A_{ij}$), and bond potential parameters $ k=20,
	r_0=1$. 
	\rev{From the balance of the bonded and non-bonded forces,} 
	the resulting \rev{average} bond length is $a\approx1.09$ with a standard
	deviation of $\sigma_a = 0.22$ based on an average over $10^6$ bonds.
	\rev{The same numbers are obtained for ideal
		chains with non-bonded forces turned off, indicating that non-bonded
		interactions have no influence on the bond length distribution.
	}
	For the WCA potential \rev{of the wall where chains are grafted} we choose
	$\sigma_{WCA}=1,\epsilon=1$.  With this choice of parameters and the
	thickness of the brush, the \rev{wall to which the chains are attached} does not directly
	influence the wetting behavior of the liquid. \rev{Instead, the wetting properties are purely
		determined by the interactions of the droplet with the grafted and lubricant chains}. All simulations are
	performed in the absence of any gravitational forces.  The time-step
	of the simulation was chosen $~\Delta t=10^{-3}$.
	
	\begin{table}[h]
		\centering
		\captionsetup{width=\linewidth}
		\begin{tabular}{ |c|c|c|c|c|c|c|c|c|c|c|c| } 
			\hline
			& $A_{pp}$ & $A_{ll}$ & $A_{pl}$\\
			\hline
			W-S & -28 & -50 & -21\\
			\hline
			W-H & -32.5 & -53 & -20\\
			\hline
			D-H & -32.5 & -50 & -21.5\\
			\hline
		\end{tabular}
		\caption{Values for the 
			\rev{parameters $A_{ij}$}
			of the attractive DPD force for the different systems we simulate. A subscript `$p$' corresponds to polymer, while `$l$' corresponds to liquid, and two letters correspond to the interaction between the two species.}
		\label{tab:interactionStrength}
	\end{table}
	
\begin{figure}[!t]
		\begin{center}
			\includegraphics[width=7cm]{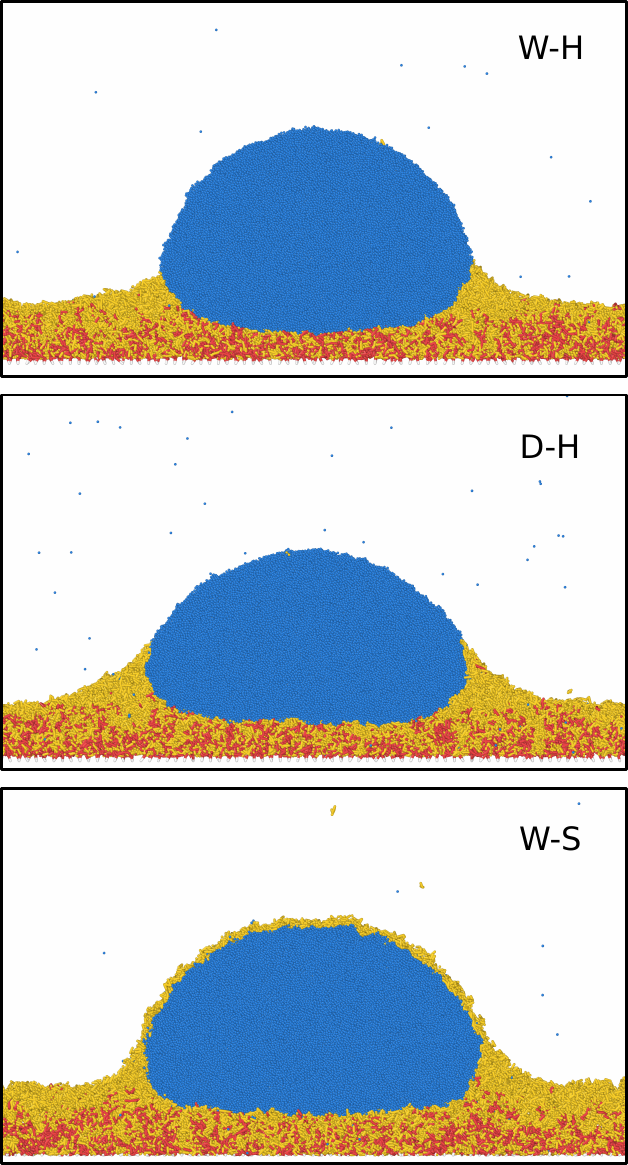}
		\end{center}
		\caption{Snapshots from simulations of the systems analogous to
			water on PLMA, DMSO40 on PLMA, and water on PDMS. Yellow chains
			are oil, red are grafted chains, and blue particles constitute the
			liquid. \rev{The snapshots for the W-H and D-H systems do not correspond to the equilibrium configurations.} }
		\label{fig:snapshots_simulations}
	\end{figure}

	To produce systems mimicking the experimental systems, we chose the
	interaction parameters such that the surface tensions between the
	different phases in simulation reproduce the relevant contact angles,
	i.e. contact angle of the liquid drop on a bulk of the brush material
	under ambient conditions. This sets our choices of the DPD parameters
	for the polymer-polymer cohesion $A_{pp}$, the liquid-liquid cohesion
	$A_{ll}$, and the polymer-liquid adhesion $A_{pl}$. Since we have
	three different combinations of drop and brush, we introduce a
	shorthand notation to refer to each combination. We use W-S
	(water-silicone) to refer to the water on PDMS and silicone oil, W-H
	(water-hexadecane) to refer to water on PLMA and hexadecane, and D-H
	(DMSO40-hexadecane) to refer to DMSO40 on PLMA and hexadecane. Table
	\ref{tab:interactionStrength} shows the final values of the $A_{ij}$
	parameters for the different systems. Our choice for the W-S system is
	motivated by our choice in previous work
	\cite{badr2022cloaking,badr2024dynamics}. As for the W-H system, to
	match the contact angles, we increased both the liquid-liquid cohesion
	strength $A_{ll}$ and the polymer-polymer cohesion $A_{pp}$, although
	the drop is still supposed to represent water. The other option was to
	reduce $A_{pp}$ while keeping $A_{ll}$ constant; however, this lead to
	a much larger vapor pressure for the oil and destabilized the system.
	Therefore, we opted to increase $A_{pp}$ and used the same value for
	the W-H and D-H systems, while tuning the values of $A_{ll}$ and
	$A_{pl}$ to match the contact angles. Each choice for the set of
	attraction strengths determines the densities, interfacial
	tensions, and viscosities of the different species and interfaces.
	Table \ref{tab:simulationDataTable} summarizes the resulting values of
	the physical quantities for each system of interest, and Fig.\ 
	\ref{fig:snapshots_simulations} shows some representative snapshots. The surface tension and viscosity are calculated as described in Refs. \citenum{badr2022cloaking} and \citenum{badr2024dynamics} (SI) respectively.
	
	The systems for simulation are prepared as described in section
	\ref{sec:systemPrep}. The simulations are conducted using the
	HOOMD-Blue simulation package
	\citep{anderson2020hoomd,phillips2011pseudo} version 4.5.0. All
	snapshot visualizations are made with the OVITO visualization package
	\citep{stukowski2009visualization}. Details of certain analysis
	methods are described in section \ref{sec:methodsSI}.

	\subsection{Experiments}
	
	Experiments were carried out using polymer brushes grown by
	surface-initiated activators regenerated by electron transfer radical
	polymerisation (SI-ARGET-ATRP) on microscope cover slips following the
	method described in refs.
	\cite{kap2023nonequilibrium,Siekmann_2026methods}. The dry thickness
	of the brushes is $160\pm5 \rm {nm}$. Brushes were impregnated with a
	fluorescent dye (nile red) and pre-swollen by exposing them to
	hexadecane vapor for up to 7 minutes, leading to a variable degree of
	swelling with swelling ratios in the range $\alpha=1 ... 4$. The
	actual experiments were initiated by depositing a millimeter-sized
	drop of a 40-60\% weight/weight DMSO-water onto the pre-swollen brush
	layer. Drop deposition was followed by a quick spreading process
	within $\sim 1\rm s$ to a steady drop shape with a macroscopic contact
	angle of $\approx 90^{\circ}$. More details will be reported elsewhere
	\cite{Siekmann_Liu_2026_brush_profiles}.

	\subsection{Continuum model} \label{sec:diffusionModelMethods}
	
	To describe the dynamics of oil transport, we construct a continuum
	model based on the free energy of the brush and its coupling to the
	contact line. The model is formulated in terms of the local volume
	fraction of oil, $\phi(\mbf{r})$. 
	We begin by defining the free energy of the brush as a functional of 
	$\phi(\mbf{r})$, 
	\begin{equation}
		\frac{\mathcal{F}_B[\phi(\mbf{r})]}{k_B T} = \int \ud V \, \frac{f_B(\phi(\mbf{r}))}{\rev{k_B T}},
	\end{equation}
	where $f_B(\phi(\mbf{r}))$ is a free energy density. Assuming
	azimuthal symmetry and fast vertical equilibration compared to lateral
	transport, we take the concentration to depend only on the radial
	coordinate $\rho$. Under this assumption, the free energy density is
	written as
	\begin{equation}
		\label{eq:freeEnergyDensity_brush}
		\frac{f_B(\phi(\rho))}{\rev{k_B T}} = \frac{\phi}{N_o \rev{a^3}} \ln{\phi} +
		\frac{\chi(\phi)}{N_o \rev{a^3}} \phi \lrp{1-\phi} + \frac{k}{2}
		\frac{\sigma^2 \rev{a}}{1-\phi},
	\end{equation}
	where $N_o$ is the number of monomers per oil chain, \rev{$a$ is the size of a monomer}, $\chi(\phi)$ is
	a \rev{dimensionless} interaction function, $k$ is the \rev{dimensionless} elastic constant of the brush, and
	$\sigma$ is the grafting density of the brush \rev{with unit $[L]^{-2}$}. The first term in Eq.~
	\ref{eq:freeEnergyDensity_brush} represents the translational entropy
	of the oil molecules, the second term accounts for interactions
	between oil and grafted chains through a concentration-dependent
	interaction parameter $\chi(\phi)$
	\cite{baulin2003signatures,schubotz2025positive}, and the third term
	captures the elastic penalty associated with stretching the grafted
	polymers. The function $\chi(\phi)$ is not known {\em a priori} but
	can be determined following an approach similar to that of Schubotz {\em et
		al.} \cite{schubotz2025positive}; details of the calculation for our
	simulated brushes can be found in Sec. \ref{sec:variableChiCalc}. \rev{Most importantly, when the brush is saturated, we expect the lubricant to act as an athermal solvent since we have $N_B > N_o^2$ \cite{rubinstein2003polymer}. Therefore, we construct our interaction function to vanish when the brush is saturated, so that we recover the free energy of a brush with an athermal solvent, namely $\chi=0$.}
	
	In addition to the free energy of the brush, we include the contribution
	of the three phase contact line through a free energy
	line-density $f_\rm{cl}(\lambda)$ where $\lambda$ is the line density of
	oil at the three phase contact line. The corresponding contribution to
	the total free energy $\mathcal{F}_\rm{cl}$ is 
	\begin{equation}
		\frac{\mathcal{F}_\rm{cl} [\lambda]}{\rev{k_B T}} = \oint \ud l \,
		\frac{f_\rm{cl}(\lambda)}{\rev{k_B T}},
	\end{equation}
	where the integral is evaluated along the three-phase contact line. 
	We assume a quadratic form for the line free energy density
	\begin{equation}
		\label{eq:freeEnergyDensity_contactLine}
		\frac{f_\rm{cl}(\lambda)}{\rev{k_B T}} = \frac{\kappa}{2} \lrp{\lambda - \lambda_0}^2,
	\end{equation}
	where $\kappa$ is an effective stiffness, and $\lambda_0$ is the
	saturation line density. \rev{In the following, we set $k_B T=1$ and $a=1$.} The dynamics in the system is driven by the
	difference in chemical potential between the contact line and the
	brush. At the contact line, we assume that the rate of change of the
	line density $\lambda$ is proportional to the chemical potential
	difference and to the local oil availability. This leads to the
	evolution equation
	\begin{equation}
		\label{eq:evolutionEq_lambda1}
		\partDer{\lambda}{t} = 
		-\mathcal{B} \Phi(\rho=R_{\rm{cl}}) \lrb{f_\rm{cl}'(\lambda) - \mu_B(t)},
	\end{equation}
	where $\mathcal{B}$ is a filling rate, $\Phi(\rho) = \int \phi (\rho)
	\, \ud z$, and $R_{\rm{cl}}$ is the radial position of the contact
	line.  To simplify the picture, we further assume that the chemical
	potential in the brush is always close to its equilibrium value
	$\mu_B^{\rm{eq}}$. This allows to rewrite the evolution equation
	for $\lambda$, Eq.\ \ref{eq:evolutionEq_lambda1}, as
	\begin{equation}
		\label{eq:evolutionEq_lambdaFinal}
		\partDer{\lambda}{t} 
		= -\mathcal{B} \Phi(R_{\rm{cl}}) \kappa \lrp{\lambda - \lambda_e}
	\end{equation}
	with the equilibrium line density
	\begin{equation}
		\label{eq:lambda_equil}
		\lambda_e
		= \lambda_0 + \frac{\mu_B^{\rm{eq}}}{\kappa},
	\end{equation}
	which depends on the equilibrium oil fraction $\phi^{\rm{eq}}$ in the
	brush. 
	
	To describe transport within the brush, we consider radial currents
	on either side of the contact line at $\rho = R_{\rm{cl}}$. The
	radial flux is expressed as
	\begin{equation}
		\label{eq:diffusionCurrent_def}
		j_\rho = -M \phi \nabla_\rho 
		\frac{\delta \mathcal{F}_B \lrb{\phi}}{\delta \phi},
	\end{equation}
	where $M$ is the mobility of oil in the brush, $\nabla_\rho$ is the
	radial component of the gradient operator in cylindrical coordinates,
	and $\delta / \delta \phi$ denotes the variational derivative with
	respect to the concentration profile $\phi(\rho,z)$. The evolution
	of the concentration field then follows from the continuity equation
	\begin{equation}
		\partDer{\phi}{t} = - \nabla_\rho j_\rho.
	\end{equation}
	The full equations are presented in full detail in the SI. 
	
	To solve the equations, appropriate boundary are imposed.
	At the origin ($\rho=0$), symmetry requires
	\begin{equation}
		\label{eq:boundaryCondition_origin}
		j_\rho(0,t) = 0.
	\end{equation}
	At the outer boundary layer, the brush is assumed to be in contact
	with a reservoir that fixes the oil fraction,
	\begin{equation}
		\label{eq:boundaryCondition_edge}
		\phi(L,t) = \phi_B.
	\end{equation}
	At the contact line ($\rho = R_{\rm{cl}}$), conservation of mass
	requires the rate of change of the line density to equal the net flux
	from both sides
	\begin{equation}
		\label{eq:boundaryCondition_contactLine}
		\partDer{\lambda}{t} = - 
		H(\phi(R_{\rm{cl}})) \lrp{j_\rho^{>}(R_{\rm{cl}}) - j_\rho^{<}(R_{\rm{cl}})}
	\end{equation}
	where $H(\phi(R_{\rm{cl}}))$ is the brush height at the contact line
	and $j_\rho^{<}(R_{\rm{cl}})$, $j_\rho^{>}(R_{\rm{cl}})$ refers to
	the fluxes at the inner and outer side of the contact line,
	respectively.
	
	For numerical implementation, the radial domain is discretized into
	bins of width $\Delta r$, with positions $r_i$.  Spatial derivatives
	are approximated using finite differences,
	\begin{align}
		&\frac{\partial g_i}{\partial r} = \frac{1}{2 \Delta r} ( g_{i+1} - g_{i-1} )\\
		&\frac{\partial^2 g_i}{\partial r^2} = \frac{1}{\Delta r^2} ( g_{i+1} + g_{i-1} - 2 g_{i} )
	\end{align}
	and time integration is performed using a forward Euler scheme.
	The boundary conditions are implemented numerically by
	enforcing
	\begin{equation}
		\label{eq:boundaryCondition_num}
		\phi_0(t) = \phi_1(t), \qquad
		\phi_L(t) = \phi(L,t) = \phi_B.
	\end{equation}
	
	\rev{To impose the conservation of mass at the contact line, we discretize Eq. \ref{eq:boundaryCondition_contactLine} using forward and backward expressions for finite derivatives. This results in an algebraic equation that we can solve for the fraction of oil at the contact line $\phi(R_{\rm{cl}})$. The full form of the equation we need to solve is given in Eq. \ref{eq:boundaryCondition_contactLine_num}.} The model contains several parameters that can be extracted from
	simulations, including the elastic constant $k$, the interaction
	function $\chi(\phi)$, and the equilibrium line density $\lambda_e$.
	The procedures to determine these parameters is described in Sec.\
	\ref{sec:modelParametersCalc}.  The mobility $M$ and the kinetic
	coefficient $\mathcal{B}$ are treated as fitting parameters and
	adjusted to match the simulation results.
	
	
	\section{Results and Discussion}
	
	
	\subsection{Kinetics of ridge and cloak development}
	
	We first examine the kinetics wetting ridge growth and cloaking.
	Ridge growth is quantified by the evolution of the height of the ridge
	above the unperturbed brush. For water on Hexadecane (W-H), we also
	analyze oil separation within the ridge. For water-PDMS systems (W-S),
	which exhibit clear signs of cloaking, we focus on the propagation of
	the cloak front and the thickening of the cloaking layer.
	
	\subsubsection{DMSO40 on PLMA}
	
	\begin{figure*}[!t]
		\begin{center}
			\includegraphics[width=14cm]{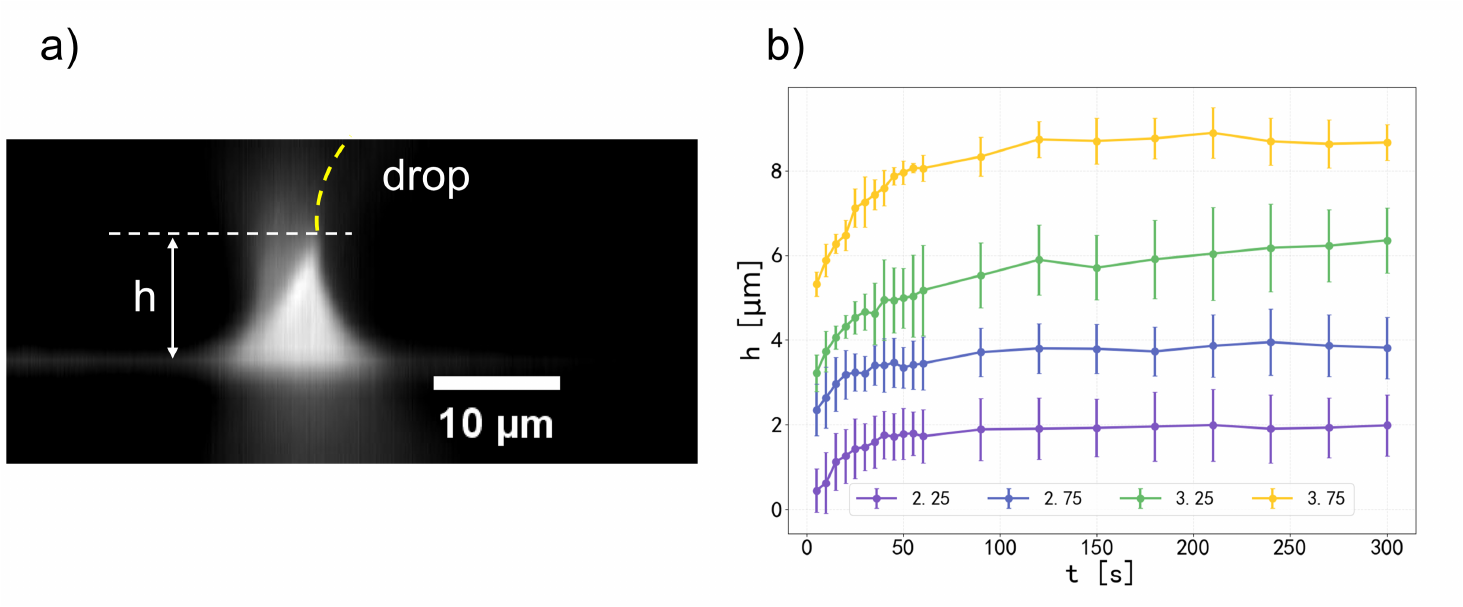}
		\end{center}
		\caption{(a) Confocal fluorescence microscopy image of a wetting
			ridge for a DMSO40 drop on a PLMA brush swollen with hexadecane
			after 300s; swelling ratio: $\alpha=3.25$. $h$ indicates of apex
			of the wedge above the \rev{brush}. (b) Evolution of the height of
			the ridge with time for brushes with different swelling ratios
			(see legend). $t=0$ indicates the time of the first snapshot after
			drop deposition and spreading, which involves some wedge growth
			for higher degrees of initial swelling.}
		\label{fig:heightVtime_PLMA_experimental}
	\end{figure*}
	
	We begin by discussing the results for DMSO40 on PLMA brushes, swollen
	with hexadecane. The spreading parameter of the oil on the drop in
	this case is positive $S_{p/l} = 2.1 \pm 1.1 \, \rm{mN/m}$. Therefore,
	a cloaking layer could form on the drop. Figure
	\ref{fig:heightVtime_PLMA_experimental} shows the experimental results
	for a growing oil wedge for this system. Figure
	\ref{fig:heightVtime_PLMA_experimental} (a) shows a typical confocal
	microscopy image of the oil forming a wetting ridge around the DMSO40
	drop for a brush with a swelling ratio of $\alpha=3.25$. The ridge
	appears as a triangular fluorescent wedge with the drop on its right
	side; the height is measured as the distance between the surface of
	the brush and the apex of the wedge. Figure
	\ref{fig:heightVtime_PLMA_experimental} (b) shows the evolution of the
	height of the ridge with time, where the error bars are standard
	errors over 4 repetitions of the experiment. The height grows
	gradually with time and seems to saturate at a maximum height that
	increases with the swelling ratio. Despite the positive spreading
	parameter of hexadecane on the DMSO40 solution, we do not see signs of
	cloaking. This could either indicate that the saturation level of the 
	brush is still below the cloaking transition \cite{badr2022cloaking},
	or that the microscope cannot resolve the presumably very thin cloaking
	layer. However, the experiments show signs of separation of the oil
	from the brush, as the maximum height of the wedge exceeds the maximum
	height of the fully swollen brush by more than a factor of $10$ for
	the most swollen brushes. Given the positive spreading parameter of
	the oil on the DMSO40 solution this is not surprising; it is
	nonetheless an interesting observation since the separation may play a
	role in the depletion of oil from the system. 
	
\begin{figure}[hb]
		\begin{center}
			\includegraphics[width=7cm]{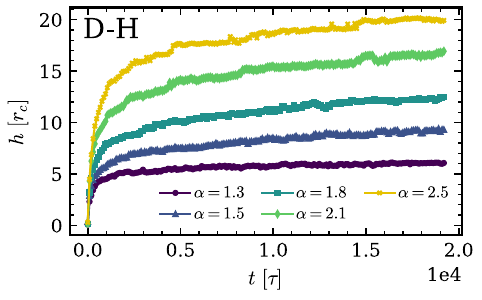}
			\includegraphics[width=7cm]{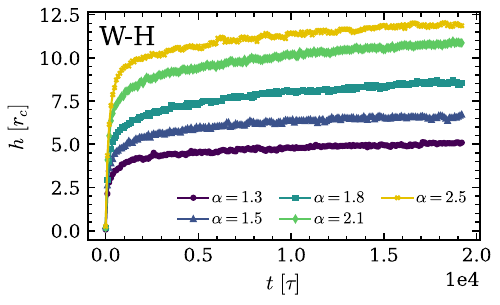}
		\end{center}
		\caption{Height above the brush reached by the apex of the ridge versus time for the D-H and W-H systems at different swelling ratios.}
		\label{fig:heightVtime_PLMA}
	\end{figure}

	To gain more insight on the molecular scales, and investigate the
	possible reason for the absence of a cloak, we run molecular dynamics
	(MD) simulations of a system mimicking the DMSO40 drop on a PLMA brush
	swollen with hexadecane (D-H). As described above, the interaction
	parameters are chosen to match the experimental contact angle and the
	sign of the spreading parameter. It should be noted, however, that the
	swelling ratios accessible in simulation are lower than those in
	experiments, owing to the significantly shorter chain lengths employed
	in the coarse-grained model. \rev{The maximum swelling ratio for our PLMA-like brushes is $\alpha^*\approx2.5$.}  The ridge height in simulations is
	determined analogously to the experimental procedure (see Sec.
	\ref{sec:densityContour}). The temporal evolution of the ridge height
	for the D–H system is shown in Fig. \ref{fig:heightVtime_PLMA}(a). The
	results indicate that the system has not reached equilibrium within
	the accessible simulation times, except possibly at the lowest
	swelling ratio. Notably, even for an oversaturated brush with \rev{$\alpha
		= \alpha^*$}, no cloaking of the DMSO40 drop is observed within the
	simulation window. This suggests that cloaking is kinetically hindered
	and that the system has not evolved for a sufficiently long time to
	reach its equilibrium configuration. \rev{The combined effect of the brush chains being longer than the lubricant chains and the fact that they are grafted can lead to an effectively more viscous environment in the brush. This in turn can result in slower diffusion of the lubricant in the brush when compared to a medium of pure lubricant. In fact, experimental investigations of the diffusion of hexadecane in PLMA brushes show that it is much slower compared to pure oil \cite{Siekmann_Liu_2026_brush_profiles}.}
	
	To test this hypothesis, we perform additional simulations in which
	oil droplets are brought into contact with liquid droplets in the
	absence of grafted chains, thereby removing the constraints imposed by
	the brush. The final configurations after a simulation time of $6.4
	\times 10^3,\tau$ (Fig. \ref{fig:snapshot_contactingDrops_SI}) clearly
	show that, for the D–H system with a positive spreading parameter
	($S_{p/l} > 0$), the oil fully engulfs the liquid, resulting in a
	core–shell morphology. This observation confirms that cloaking is
	thermodynamically favorable in this system. We therefore expect that,
	in the presence of the brush, a fully cloaked state would eventually
	be reached at sufficiently large times and swelling ratios. However,
	accessing these regimes remains computationally prohibitive within the
	present study.
	
	\subsubsection{Water on PLMA}

	\begin{figure*}[ht]
		\begin{center}
			\includegraphics[width=7cm]{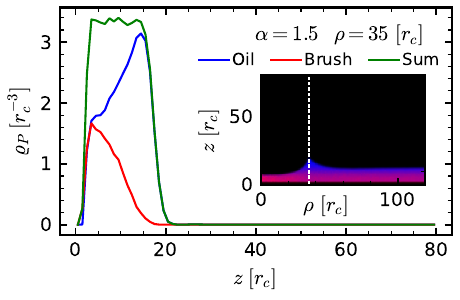}
			\includegraphics[width=7cm]{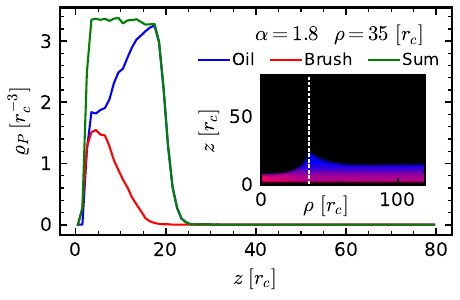}
		\end{center}
		\caption{Vertical density profile of polymers in the W-H system in
			the area of the wetting ridge at radial position $\rho=35 \,
			[r_c]$ for swelling ratios $\alpha=1.5$ and $\alpha=1.8$. Insets
			show the density map $\varrho(\rho,z)$ with the grafted chains in
			red and the oil in blue. Magenta indicates a mixture of the two. 
			\rev{Although the insets suggest that far from the droplet, the brush is covered by a layer of oil, the profiles show that is not the case -- the change of color merely reflects the accumulation of oil in the polymer/vapor interface without the formation of a bulk oil layer (see Figure \ref{fig:fluidSeparation_PLMA_SI_far}).}
			The dashed lines indicate the position where the density profile
			is shown. For the higher swelling ratio we clearly see that the
			fluid separates from the brush as indicated by a density of oil
			equal to the bulk density of polymers. }
		\label{fig:fluidSeparation_PLMA}
	\end{figure*}
	
	Since water is a prevalent liquid in many applications, it is of
	particular interest to investigate the interaction of water drops with
	lubricated polymer brushes. Experimentally, however, this system is
	difficult to probe using top-view microscopy, as the large contact
	angle of water on PLMA (exceeding $100~\circ$) obscures direct
	observation of the wetting ridge. We therefore restrict our analysis
	of this system to MD simulations.
	
	Fig. \ref{fig:heightVtime_PLMA}(b) shows the time evolution of the
	ridge height for systems corresponding to water on a PLMA brush
	swollen with hexadecane (W–H) at different swelling ratios. As in the
	D–H system, the ridge height increases with time and does not fully
	saturate within the accessible simulation window, except at the lowest
	swelling ratio. However, at comparable swelling, the ridge height in
	the W–H system is consistently smaller than in the D–H case. Since the
	spreading parameter of oil on water in the W-H system is negative
	($S=-0.43 \pm 0.06 \, [k_BT/r_c^2]$), cloaking is not expected. \rev{In this case, we expect the oil and liquid to adopt a Neumann configuration for the angles (see section \ref{sec:NeumannAngles}). Based on our values for the surface tension in simulation, we expect a triplet of angles $(161.42^\circ,143.01^\circ,55.57^\circ)$ . Since the simulations with the brush did not reach equilibrium, we calculate the contact angles between oil and liquid from contacting droplet simulations (see Figure \ref{fig:snapshot_contactingDrops_SI}). The values of the angles calculated from those simulations are $(164.06^\circ,137.82^\circ,58.12^\circ)$. The values from the simulations agree reasonably with the values obtained from the Neumann balance.}
	Despite the absence of cloaking, inspection of simulation snapshots (see Fig.
	\ref{fig:snapshots_simulations}) suggests that the oil can partially
	separate from the brush within the wetting ridge. Since such
	separation may enhance lubricant depletion, it is important to
	characterize this phenomenon in more detail.
	
	Oil separation has previously been reported and quantified for water
	drops on PDMS elastomers swollen with silicone oil \cite{cai2021fluid,
		cai2024phase}, where the spreading parameter is positive and cloaking
	can occur. In that context, separation is a necessary precursor to
	cloaking, although it has also been observed in cases where no visible
	cloak forms \cite{cai2021fluid, cai2024phase}.
	In contrast, for systems with a negative spreading
	parameter, such as hexadecane on water, the occurrence of oil
	separation is less straightforward and requires careful analysis.
	
	To address this, we examine the final configurations of the W-H system
	at different swelling ratios. \rev{The maximum swelling ratio for the PDMS-like brush is $\alpha^*\approx2.3$} Fig. \ref{fig:fluidSeparation_PLMA}
	shows the vertical density profiles $\varrho(\rho_{\rm{ridge}},z)$ of
	grafted chains, oil, and their sum at the position of the wetting
	ridge, (radial coordinate $\rho_{\rm{ridge}} = 35 \, [r_c]$, see white
	dashed line in the inset), for swelling ratios $\alpha=1.5$ and
	$\alpha=1.8$.  The insets display the full density map of polymers
	$\varrho_P(\rho,z)$, where grafted chains are shown in red and oil in
	blue; mixed regions appear in magenta. At the lower swelling ratio
	($\alpha=1.5$), the oil remains well mixed with the brush throughout
	the ridge region, as indicated by the oil density remaining below the
	total polymer density. In contrast, at the higher swelling ratio
	($\alpha = 1.8$), the oil density near the top of the ridge approaches
	the total density, indicating the onset of phase separation.
	Additional data for other swelling ratios (see Fig.
	\ref{fig:fluidSeparation_PLMA_SI}) confirm this trend. Moreover,
	density profiles taken far from the drop ($\rho=80$,
	\ref{fig:fluidSeparation_PLMA_SI_far}) show that, for
	undersaturated brushes, the oil remains fully incorporated within the
	brush. \rev{This observation also informs the proper reading of the colored density maps in the insets of Figures \ref{fig:fluidSeparation_PLMA}, \ref{fig:fluidSeparation_PLMA_SI}, and \ref{fig:fluidSeparation_PLMA_SI_far}. One may be tempted to see a blue color and conclude that the oil separated from the brush. However, this is an optical effect, and the true nature of the separation can only be concluded from the density profiles.}
	
	These results demonstrate that the presence of a water drop can induce
	local separation of oil from the brush within the wetting ridge, even
	in the absence of cloaking, which in turn affects the rate of oil
	depletion in applications. This separation only occurs after the brush
	is sufficiently swollen.

	\subsubsection{Water on PDMS}
	
	When the polymer brush consists of PDMS swollen with 
	chemically identical silicone oil, the spreading parameter of the oil
	on a water drop is significantly positive ($S_{p/l} \approx 12
	\rm{mN/m}$). Under these conditions, cloaking of the drop sets in
	once the brush is sufficiently swollen~ \cite{badr2022cloaking}. We
	therefore focus on quantifying the kinetics of cloak formation.
	To this end, we first track the evolution of the cloaking
	layer by following the position of the cloak front,
	which gradually advances toward the apex of the droplet.
	
	To find the position of the cloak front, we calculate the azimuthally
	symmetric local density of oil $\varrho_o(r,\theta)$ in spherical
	coordinates as described in section \ref{sec:densityMapsCalc} (see
	Fig. \ref{fig:densityMaps_SI} b). From the density map, we find the
	equal density contours $\varrho_o(r,\theta) = \rho_o/2$. Afterwards,
	we calculate the angular position reached by the cloak as the smallest
	angular value on the contour. The arc length covered by the cloak can
	then be calculated as
	\begin{equation}
		s(t)= (\theta_{\rm{app}}(t)-\theta_{\rm{front}}(t)) \times R(t),
	\end{equation}
	where $\theta_{\rm{app}}$ is the apparent contact angle of the drop,
	$\theta_{\rm{front}}$ is the polar angle reached by the cloak front,
	and $R_{\rm{D}}$ is the radius of curvature of the drop.

	\begin{figure}[t]
			\begin{center}
				\includegraphics[width=7cm]{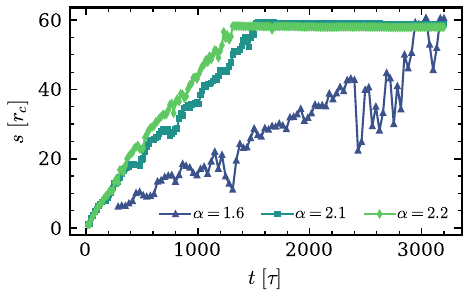}
			\end{center}
			\caption[Cloaking front position versus time]{Cloak front position
				versus time for three different swelling ratios $\alpha$, all
				beyond the cloaking transition. As expected the higher the
				swelling ratio, the faster the cloaking progresses.}
			\label{fig:cloakFrontProgress}
		\end{figure}
		
		The temporal evolution of the cloak front is shown in Fig.
		\ref{fig:cloakFrontProgress} for several swelling ratios above the
		cloaking threshold. In all cases, the cloak front advances
		approximately linearly in time, indicating a nearly constant
		propagation speed. Moreover, the speed increases with increasing
		swelling ratio.
		
		\begin{figure}[h]
				\begin{center}
					\includegraphics[width=7cm]{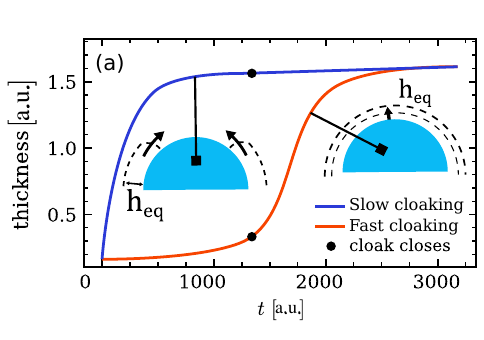}
					\includegraphics[width=7cm]{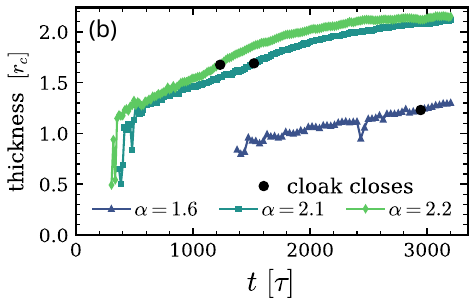}
				\end{center}
				\caption{(a) Qualitative representation of the two extreme
					thickening regimes of the cloak. (b) Thickness of the cloak versus
					time for the three swelling ratios we studied. In both figures the
					black dots refer to the time and thickness when the cloak reaches
					the top of the droplet.}
				\label{fig:cloakThickVsTime}
			\end{figure}
			
			Beyond the position of the front, an important aspect of the dynamics
			is the evolution of the cloak thickness. Two limiting regimes can be
			identified. In the ``fast cloaking'' regime, transport along the drop
			surface is rapid compared to transport within the brush, leading to a
			thin layer that quickly covers the entire drop before thickening. In
			the opposite ``slow cloaking'' regime, transport along the surface is
			rate-limiting, such that the cloak locally reaches its equilibrium
			thickness before advancing further. Fig. \ref{fig:cloakThickVsTime}
			(a) shows qualitative expectations for the time evolution of the
			thickness of the cloak in each of the two extremes.  In the fast
			cloaking regime, we expect the thickness to remain small and
			approximately constant during spreading, and increase only after the
			cloak reaches the apex of the drop. In contrast, in the slow cloaking
			regime, the thickness should grow rapidly at early times and
			subsequently saturate once its local equilibrium value is reached.
			
			To determine which regime applies in our system, we compute the time
			evolution of the cloak thickness as described in Sec.\
			\ref{sec:cloakThicknessCalc}. The results are shown in Fig.
			\ref{fig:cloakThickVsTime} (b) for different swelling ratios. The
			black markers indicate the time at which the cloak front reaches the
			apex of the drop, defined as the point where the front lies within
			$3^\circ$ of the top.  The observed behavior does not conform to
			either limiting case. Instead, the cloak thickness increases
			continuously both before and after the front reaches the apex,
			indicating that neither surface transport nor diffusion within the
			brush is overwhelmingly dominant. This places the system in an
			intermediate regime, where the characteristic timescales for transport
			along the drop surface and within the brush are comparable. Such a
			balance of timescales leads to simultaneous spreading and thickening
			of the cloaking layer.
			
			
			\subsection{Material transport during growth}
			
			Having characterized the kinetics of wetting ridge growth and
			cloaking, we now turn to the underlying material transport within the
			brush. In particular, we focus on how oil is redistributed during the
			approach to equilibrium and how this redistribution governs the
			observed dynamics. Our simulation results motivate the development of
			a theoretical framework for oil transport that captures the kinetics
			of ridge growth.

			\subsubsection{Oil depletion}
			
			The results of the previous section demonstrate that the growth rate
			of both the wetting ridge and the cloaking layer depends on the
			swelling ratio of the brush. This dependence reflects the local
			availability of oil near the contact line. As the ridge grows, it
			continuously draws in material from its surroundings, leading to
			depletion of oil in adjacent regions. Consequently, further growth
			requires transport of oil from increasingly distant parts of the
			brush. To quantify this redistribution, we introduce the integrated
			oil density
			\begin{equation}
				\Lambda(\rho) = \int_{\rev{0}}^{\rev{L_z}} \varrho_{p}(\rho, z) dz,
				\label{eq:lambda}
			\end{equation}
			which represents an effective area density of oil at radial position 
			$\rho$, averaged over the azimuthal direction.

			\begin{figure}[t]
					\begin{center}
						\includegraphics[width=7cm]{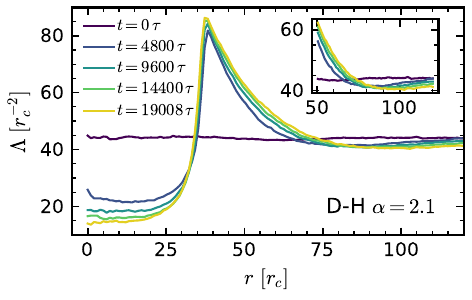}
						\includegraphics[width=7cm]{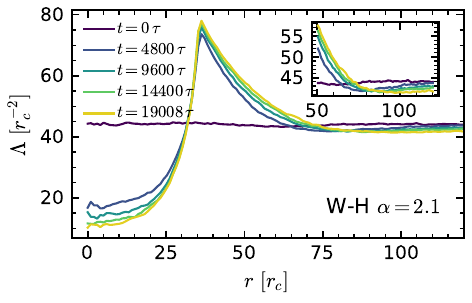}
						\includegraphics[width=7cm]{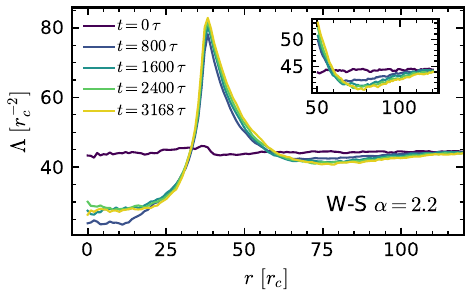}
					\end{center}
					\caption{$\Lambda(\rho)$ at different time points during cloaking
						for the D-H, W-H, and W-S systems. Insets are close ups near the
						three phase contact line. There is a clear depletion zone of
						material outside of the drop. The amount of material also drops
						from under the droplet.}
					\label{fig:LambdaVsTime}
				\end{figure}
				
				Fig. \ref{fig:LambdaVsTime} shows the radial profiles
				$\Lambda(\rho)$ for the different systems at several time points
				during the simulation; the insets provide a magnified view of the
				region outside of the drop.  Initially, the profile is approximately
				uniform, reflecting the homogeneous state of the unperturbed brush. As
				the system evolves, the presence of the drop induces the growth of the
				wetting ridge in the vicinity of the contact line, which is at a
				radius $35 \lesssim R_{\rm{cl}} \lesssim 40 \, [r_c]$ depending on the
				system. \rev{For the the W-S system, the presence of the cloak will slightly affect the value of $\Lambda(r)$ where the drop is present. However, since the thickness of the cloak is much smaller than the thickness of the brush under the drop (see our previous work Ref. \citenum{badr2022cloaking}), we do not expect this contribution to affect the conclusions we draw from our $\Lambda(r)$ profiles. }
				
				Beneath the drop ($r < R_{\rm{cl}}$), a pronounced decrease in
				$\Lambda(\rho)$ is observed. This reduction arises from two distinct
				mechanisms. First, oil is depleted as it is transported toward the
				contact line and incorporated into the growing ridge. Second, the
				Laplace pressure of the drop exerts a downward stress on the brush,
				compressing the grafted chains and expelling oil from the underlying
				region. At the contact line ($r \approx R_{\rm{cl}})$) $\Lambda(\rho)$
				exhibits a pronounced peak, corresponding to the accumulation of oil
				in the wetting ridge. Moving further outward ($r>R_{\rm{cl}}$), a
				shallow minimum develops, indicating the formation of a depletion zone
				for polymers outside the drop.  At low swelling ratios, this depletion
				remains localized near the ridge (see Figure
				\ref{fig:LambdaVsTime_lowSwelling}). For large swelling ratios, the
				depletion zone extends over substantial distances and in some cases
				reaches the boundaries of the simulation domain. Ideally, one should
				simulate larger boxes; however, this was not possible due to
				computational limitations.

				\subsubsection{Modeling the material transport}

				\rev{The presence of depletion zones both beneath and outside the drop
					indicates that oil is removed from the brush faster than it can be
					replenished locally. As a result, continued growth of the ridge relies
					on the transport of oil from distant regions. This observation
					highlights the importance of long-range diffusion within the brush and
					suggests that the overall kinetics are limited by the rate at which
					oil can be supplied to the contact line.  These findings motivate the
					development of a continuum description of the transport process. We
					adapt a diffusion-based model introduced in previous work
					\cite{cai2024phase} to the case of polymer brushes, and extend the
					domain to both sides of the wetting ridge, i.e. under and outside of
					the drop. In addition, we use a variable interaction function in the
					free energy that depends on the swelling of the brush.  Details of the
					model can be found in section \ref{sec:diffusionModelDerivation} with the main elements introduced in section \ref{sec:diffusionModelMethods}.}
				
				\begin{figure}[t]
						\begin{center}
							\includegraphics[width=7cm]{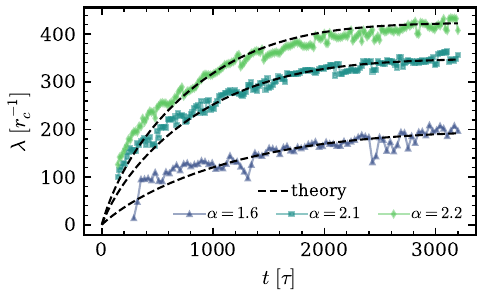}
						\end{center}
						\caption{The line density at the ridge from simulation (data points) and theory (dashed lines).}
						\label{fig:lineDensity_WS}
					\end{figure}

					To assess the validity of the \rev{continuum} model, we compare its
					predictions for the line density of oil at the contact line with the
					corresponding results from MD simulations \rev{of the W-S system}. In simulations, the line
					density is computed as
					\begin{equation}
						\lambda^{\rm{sim}}(t) = N(t)/2\pi R_{\rm{cl}},
					\end{equation}
					where $N(t)$ is s the total amount of oil contained in the ridge and
					cloak region, obtained from the density field as
					\begin{equation}
						N(t) = 2\pi R_{\rm{D}}^2 
						\int_{\theta_{\text{tip}}(t)}^{\theta_{\rm{max}}} 
						\int_{0}^{R_{\rm{max}}} \varrho_{o} \sin\theta \, \ud r \, \ud
						\theta.
					\end{equation}
					Here $R_{\rm{D}}$ is the radius of curvature of the drop,
					$R_{\rm{cl}}$ is the contact line radius, and $\theta_{\rm{max}} =
					75^\circ$.  The free parameters of the model, $M$ and $\mathcal{B}$,
					are determined by adjusting the theoretical prediction to the
					simulations at a reference swelling ratio $\alpha=2.2$. 
					This procedure yields
					\begin{equation}
						M = 10 ~~~ ; ~~~ \mathcal{B} = 0.08.
					\end{equation}
					Keeping these parameters fixed, we then vary the swelling ratio and
					compute the corresponding time evolution of the line density at
					the contact line.
					
					The comparison between model and simulation is shown in Fig.
					\ref{fig:lineDensity_WS}.  Overall, the model captures the temporal
					evolution of the line density well at intermediate and long times for
					all swelling ratios considered. In particular, it reproduces both the
					growth rate and the saturation behavior observed in the simulations.
					At early times, systematic deviations are observed, with the
					simulations exhibiting a more rapid initial increase in $\lambda$ than
					predicted by the model. This discrepancy can likely be attributed to
					advective effects associated with the initial formation of the wetting
					ridge, during which material is rapidly drawn toward the contact line.
					Such processes are not included in the present description, which
					assumes purely diffusive transport within the brush.
					
					As discussed above, the free energy functional includes a
					concentration-dependent interaction term through $\chi(\phi)$. To
					assess its impact, we also solve the model with this term omitted,
					while recalibrating the parameters $\kappa$ and $\mathcal{B}$.  The
					resulting predictions (see Fig. \ref{fig:lineDensity_WS_compare}) are
					found to be very similar to those obtained with the full model. This
					indicates that, within the parameter range explored here, the detailed
					form of the interaction term plays only a minor role in determining
					the overall kinetics.

					
					\section{Conclusion}
					
					We have investigated the kinetics of wetting ridge growth and droplet
					cloaking on lubricant-infused polymer brushes using a combined
					experimental, simulation, and theoretical approach. Our analysis
					suggests different dynamical regimes governed by the interplay between
					interfacial forces, brush elasticity, and the transport of lubricant
					within the brush.
					
					We identify distinct regimes of ridge growth and cloaking that depend
					sensitively on the swelling state of the brush and the sign of the
					spreading parameter. While a positive spreading parameter favors
					cloaking at equilibrium, our results highlight that the transient
					dynamics can be complex, with the evolution of the system controlled
					by the availability and redistribution of oil within the brush. For
					negative spreading parameter, cloaking is absent, but substantial
					restructuring of the lubricant phase may still occur near the contact
					line.
					
					A central finding of this work is the strong coupling between ridge
					growth and oil transport. The formation of the wetting ridge is
					accompanied by pronounced depletion of oil both beneath and outside
					the drop, indicating that local supply is insufficient to sustain
					growth. As a result, the evolution of the system is governed by
					transport from increasingly distant regions of the brush. Moreover
					we show that, above a critical swelling, the lubricant can locally
					separate from the brush within the ridge. This phase separation
					provides an additional pathway for lubricant depletion and is
					therefore likely to play an important role in determining the
					long-term stability of lubricant-infused surfaces.
					
					To rationalize these observations, we developed a continuum diffusion
					model based on the free energy of the brush and its coupling to the
					contact line. Despite its simplicity, the model quantitatively
					captures the evolution of the oil line density at intermediate and
					late times across a range of swelling ratios. This agreement
					indicates that the kinetics of ridge growth and cloaking are largely
					controlled by diffusive transport of oil within the brush. Deviations
					at early times can be attributed to advective processes associated
					with the initial formation of the ridge, which are not included in the
					present description.
					
					Our results also provide insight into the dynamics of cloaking on
					droplets.  In particular, we find that the propagation of the cloaking
					front and the thickening of the film can occur on comparable time
					scales, reflecting a balance between transport along the drop
					interface and within the brush. This highlights the inherently coupled
					nature of surface and bulk transport processes in these systems.
					
					Our results can be viewed in the context of previous work on
					lubricant-infused surfaces \cite{kreder2018film}, where the growth of
					the wetting ridge was identified as the primary source of lubricant
					depletion and its dynamics were rationalized in terms of pressure
					gradients and viscous stresses in a porous substrate. The present
					study extends this framework to polymer brushes, where the lubricant
					is stored within a deformable medium and transport is governed by a
					combination of diffusion and thermodynamic driving forces. In this
					setting, we further identify local phase separation within the ridge
					as an additional pathway for lubricant depletion, which is not
					captured in models of non-deformable substrates.
					
					Overall, our results demonstrate that the dynamics of wetting ridge
					growth and cloaking on polymer brushes can be understood within a
					unified framework that combines interfacial thermodynamics with
					diffusion-limited transport. These insights extend existing concepts
					for lubricant-infused surfaces to soft, deformable substrates and
					provide guidance for the design of coatings with improved resistance
					to depletion and enhanced long-term performance.
					

					
					\backmatter
					
					\bmhead{Supplementary information}
					
					Supplementary Information is available with a table containing the values of physical parameters in simulation such as surface tension and viscosity, a description of the brush free energy with varying interaction term and the method to determine that term, a derivation of the diffusion equation used to model the material transport, a description of some of the methods used in the simulations and subsequent analysis, additional data on the oil separation in the W-H system, simulation snapshots of contacting drops of liquid and oil to illustrate the presence or absence of cloaking, additional graphs for the non-equilibrium distribution of oil during ridge growth and cloaking, and a figure comparing the performance of the theoretical model with a varying interaction function to the one where interactions are absent.
					
					\bmhead{Acknowledgements}
					EL thanks Sander Reuvekamp for his help on the brush sample preparation. This work was funded by the German Science Foundation (DFG) within the priority program 
					SPP 2171 (Grant No. 422796905, projects Schm 985/22) and as part of the project "Advanced Grease
					Lubrication Based on Liquid-Infused Surfaces” (file number 19474),
					which is part of the Open Technology Program and is financed by the
					Dutch Research Council (NWO), SKF, and Shell. Further support is acknowledged from the DFG-funded Graduate School
					RTG 2516 (Grant No. 405552959): RGMB is an
					associated member, FS is a member. 
					The simulations were partly carried out
					on the supercomputer system Mogon NHR S\" ud-West at Johannes Gutenberg University Mainz.

					\bmhead{Data Availability}
					All codes used for simulation are available on the github repository \url{https://github.com/rodbadr/dropletOnLubricatedBrush-HOOMD4}. Data used for the figures along with plotting codes, as well as the codes used to solve the diffusion equation, are made available on the Zenodo repository \url{https://doi.org/10.5281/zenodo.19557302}. The exception is the data for the figures that require the density maps due to file size limitations. Such data can be made available upon reasonable request.

					\section*{Declarations}
					
					Not applicable
					

					\bibliography{refs.bib}
					
					\clearpage
					
					\onecolumn
					
					\setcounter{section}{0}
					\setcounter{figure}{0}
					\setcounter{table}{0}
					\setcounter{equation}{0}
					
					\renewcommand{\thesection}{S.\arabic{section}}
					\renewcommand{\thefigure}{S.\arabic{figure}}
					\renewcommand{\thetable}{S.\arabic{table}}
					\renewcommand{\theequation}{S.\arabic{equation}}
					

					{\Large{\textbf{Supplementary Information}}}
					
					
					\section{Simulation parameters and physical quantities}\label{sec:physicalQuantities}

					\begin{table*}[h]
						\centering
						\begin{tabular}{ |c|c|c|c|c|c|c|c|c|c|c|c| } 
							\hline
							& $\rho_{p}$ & $\rho_{l}$ & $\gamma_{p}$ & $\gamma_{l}$ & $\gamma_{pl}$ & $S_{p/l}$ & $\theta_Y$ & $\mu_p$ & $\mu_l$ \\
							\hline
							W-S & 2.9 & 4 &  $0.841(20)$ & $3.14(4)$ & $1.41(3)$ & $0.88(6)$ & $100.5(7)$ & $5.171(27)$ & $5.806(8)$ \\
							\hline
							W-H & 3.3 & 4.2 &  $1.445(14)$ & $3.74(4)$ & $2.728(20)$ & $-0.43(6)$ & $110.1(5)$ & $9.12(7)$ & $6.785(4)$ \\
							\hline
							D-H & 3.3 & 4 &  $1.49(3)$ & $3.15(3)$ & $1.54(5)$ & $0.12(4)$ & $91(1)$ & $9.12(7)$ & $5.806(8)$ \\
							\hline
						\end{tabular}
						\caption{Results for the equilibrium densities, interfacial tension values, the spreading parameter of the oil on the liquid drop $S_{p/l}$, the estimated Young contact angle, and the viscosities. W-S is the Water-Silicone oil system, W-H is Water-Hexadecane, and D-H is DMSO40-Hexadecane. A subscript `$p$' corresponds to polymer, while `$l$' corresponds to liquid, and two letters correspond to interface between the two species. All values are in simulation units set by $r_c=1~;~k_BT=1~;~m=1$. Values in parentheses are uncertainties in the rightmost digits.}
						\label{tab:simulationDataTable}
					\end{table*}
					
					
					\section{Neumann balance} \label{sec:NeumannAngles}
					
					\color{blue}
					
					\begin{figure}[h]
						\centering
						\includegraphics[width=0.8\linewidth]{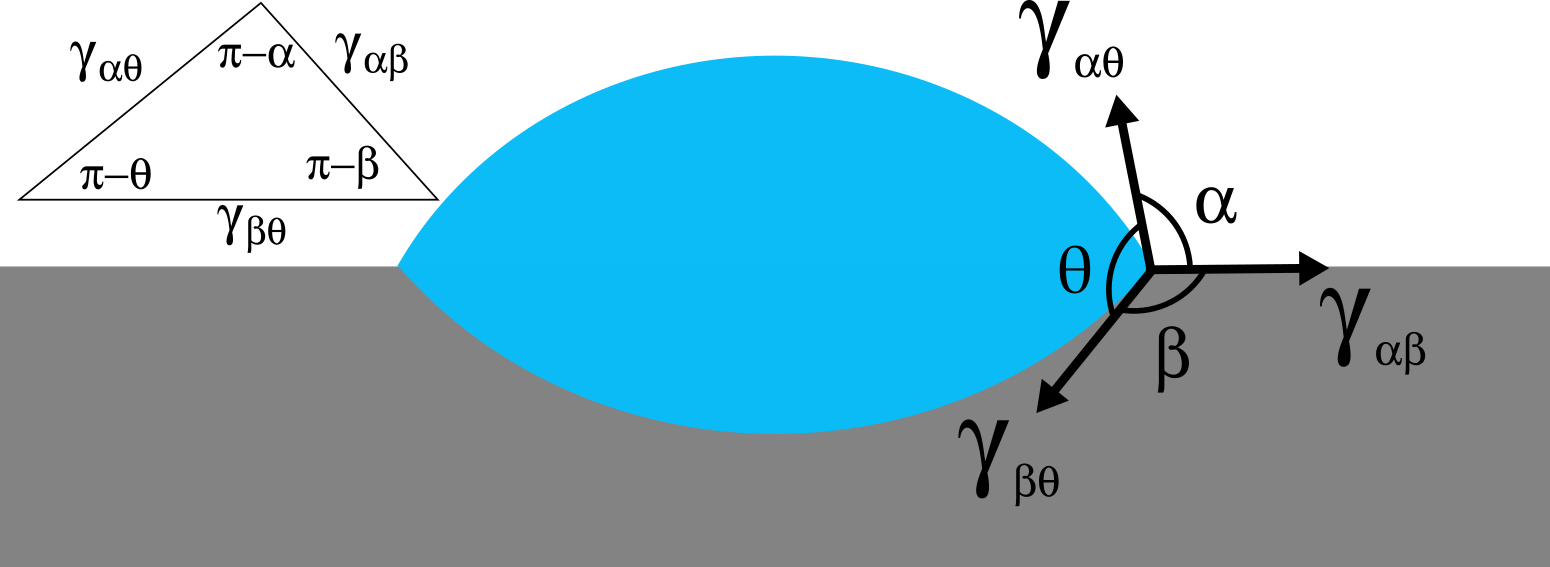}
						\caption{Cartoon of a drop on a liquid surface. The configuration is that of a liquid lens, with three angle $\alpha$, $\beta$, and $\theta$ at the three phase contact line. The angles are set by the balance of the three surface tensions between the three phases. Also shown is the corresponding Neumann triangle for this configuration. Reproduced from \cite{badr2025drops} under CC-BY-SA 4.0.}
						\label{fig:neumannIllus}
					\end{figure}
					
					For two immiscible liquids in contact, at the three phase contact line, the equilibrium is dictated by the Neumann balance of forces. We label the three fluid phases as $\alpha$, $\beta$, and $\theta$, which also corresponds to the angle in each phase. The surface tensions are labeled $\gamma_{XY}$ for the tension at the interface between phases $X$ and $Y$, with $X,~Y \in \{ \alpha$, $\beta$, $\theta \}$. The liquid surface is fully deformable and the equilibrium configuration is that of a liquid lens. Balancing forces (per unit length) in both the vertical and horizontal directions yields
					
					\begin{align}
						\nonumber
						&\gamma_{\alpha\theta} + \gamma_{\beta\theta} \cos{\theta} + \gamma_{\alpha\beta} \cos{\alpha} = 0\\ \label{eq:neumannBalance}
						&\gamma_{\alpha\theta} \cos{\theta} + \gamma_{\beta\theta} + \gamma_{\alpha\beta} \cos{\beta} = 0\\\nonumber
						&\gamma_{\alpha\theta} \cos{\alpha} + \gamma_{\beta\theta} \cos{\beta} + \gamma_{\alpha\beta} = 0.
					\end{align}
					
					The configuration satisfying the equations above is known as a Neumann configuration. Another way of identifying the angles in a Neumann configuration is through what is known as a Neumann triangle. One can construct a triangle whose side lengths are proportional to the surface tensions between the three phases as shown on the left side of Figure \ref{fig:neumannIllus}. The angles in the triangle can be mapped to the contact angles in the wetting configuration. We can see from this analogy that a Neumann configuration is only possible if the surface tensions satisfy a triangle inequality
					
					\begin{equation}
						\gamma_{XY} + \gamma_{YZ} \geq \gamma_{XZ}
					\end{equation}
					
					with $X,Y,Z \in \{\alpha,\beta,\theta\}$. The inequality needs to be satisfied for all permutations of $\alpha$, $\beta$, and $\theta$. If the triangle inequality is broken, the equilibrium configuration is either one where the deposited liquid spreads on the surface, is engulfed by it, or fully detaches from it, depending on the spreading parameters. The Neumann triangle view allows us then to write practical equations for the angles in terms of the surface tensions using the law of cosines
					
					\begin{align}
						\cos X = \frac{\gamma_{YZ}^2 - \gamma_{XY}^2 - \gamma_{XZ}^2}{2\gamma_{XY} \gamma_{XZ}}
					\end{align}
					
					where again $X,Y,Z \in \{\alpha,\beta,\theta\}$.
					
					\color{black}
					
					
					\section{Variable interaction function} \label{sec:variableChiCalc}

					
					\subsection{Free energy with constant interaction parameter} \label{sec:brushFreeEnergy_constantChi}
					
					Notation:
					
					\begin{itemize}
						\item $n_o$: number of oil molecules
						\item $N_o$: length of an oil molecule in number of monomers
						\item $n_B$: number of grafted chains
						\item $N_B$: length of a grafted chain in number of monomers
						\item $\phi$: number fraction of oil monomers in relation to grafted chain monomers
						\item $k$: elasticity of the brush assuming it deviates from the ideal value $3$
						\item $H_B$: thickness of the brush layer
						\item $\mu_B$: chemical potential in the brush
						\item $\sigma$: grafting density of the brush
						\item $a$: size of a monomer
						
					\end{itemize}

					The free energy of the brush with an arbitrary solvent can be written as \cite{badr2022cloaking, schubotz2025positive}
					
					\begin{equation}
						\label{eq:brushFreeEnergy_OG}
						\frac{\mathcal{F}}{k_B T} = n_o \ln{\phi} + \chi n_o (1-\phi) + \frac{k}{2} \frac{n_B}{N_B} H_B^2 - n_o \mu_B
					\end{equation}
					
					where the symbols are defined in the notation above. The reference free energy is that of a saturated brush so that the chemical potential vanishes at saturation. The number fraction of oil $\phi$ and that of grafted chain monomers $1-\phi$ are given by
					
					\begin{align}
						\label{eq:phi_o_def}
						\phi = \frac{n_o N_o}{n_o N_o + n_B N_B}\\ \label{eq:phi_B_def}
						1-\phi = \frac{n_B N_B}{n_o N_o + n_B N_B}
					\end{align}
					
					The height of the brush $H_B$ is given by
					
					\begin{equation}
						\label{eq:brushHeight_def}
						H_B = \frac{N_B \sigma a^3}{1-\phi}.
					\end{equation}
					
					From Eq. \ref{eq:brushFreeEnergy_OG} we write the free energy per grafted chain monomer $\Tilde{\mathcal{F}} = \mathcal{F} / (n_B N_B)$, also substituting Eq. \ref{eq:brushHeight_def} and setting $a=1$
					
					\begin{equation}
						\label{eq:brushFreeEnergy_perMon1}
						\frac{\Tilde{\mathcal{F}}}{k_B T} = \frac{n_o}{n_B N_B} \ln{\phi} + \chi \frac{n_o}{n_B N_B} (1-\phi) + \frac{k}{2} \frac{\sigma^2}{(1-\phi)^2} - \frac{n_o}{n_B N_B} \mu_B.
					\end{equation}
					
					Using Eq. \ref{eq:phi_o_def} and \ref{eq:phi_B_def} we can write
					
					\begin{align}
						\frac{n_o}{n_B N_B} &= \frac{\phi}{N_o} \frac{n_o N_o + n_B N_B}{n_B N_B}\\
						&= \frac{1}{N_o} \frac{\phi}{1-\phi}
					\end{align}
					
					which we substitute into Eq. \ref{eq:brushFreeEnergy_perMon1} to get
					
					\begin{align}
						\frac{\Tilde{\mathcal{F}}}{k_B T} &= \frac{1}{N_o} \frac{\phi}{1-\phi} \ln{\phi} + \chi \frac{1}{N_o} \phi + \frac{k}{2} \frac{\sigma^2}{(1-\phi)^2} - \frac{1}{N_o} \frac{\phi}{1-\phi} \mu_B \\ \label{eq:brushFreeEnergy_perMonFinal}
						&= \Tilde{\mathcal{F}}_0 + \chi \frac{1}{N_o} \phi
					\end{align}
					
					where $\Tilde{\mathcal{F}}_0$ includes the terms that do not depend on the interaction parameter.

					
					\subsection{Free energy with variable interaction parameter}\label{sec:brushFreeEnergy_variableChi}
					
					Following Refs. \citenum{baulin2003signatures} and \citenum{schubotz2025positive}, we can modify our free energy Eq. \ref{eq:brushFreeEnergy_perMonFinal} by making the interaction parameter depend on the fraction of oil in the brush. The equilibrium is obtained by setting $\partDerH{\Tilde{\mathcal{F}}}{\phi}=0$
					
					\begin{align}
						\label{eq:brushFreeEnergy_variableChiDer}
						\partDer{\Tilde{\mathcal{F}}}{\phi} =& \partDer{\Tilde{\mathcal{F}}_0(\phi, \mu)}{\phi} + \frac{1}{N_o} \lrb{\chi(\phi) + \phi \chi'(\phi)} = 0.
					\end{align}
					
					The differential equation can be solved for $\chi(\phi)$ and we finally get (see Ref. \citenum{schubotz2025positive} SI)
					
					\begin{equation}
						\label{eq:variableInteractionFunction}
						\chi(\phi) = \frac{N_o}{\phi} \lrp{ \Tilde{\mathcal{F}}_0(\phi^*, 0) - \Tilde{\mathcal{F}}_0(\phi, \mu)}
					\end{equation}
					
					where $\phi^*$ is the oil fraction at saturation. It is worth noting here that our Eq. \ref{eq:variableInteractionFunction} is different from that in Ref. \citenum{schubotz2025positive} as we choose the saturated brush as a reference when solving the differential equation, leading to a vanishing interaction function at saturation. \rev{The motivation for this is that the lubricant chains are much shorter than the grafted chains. Therefore, since the two are chemically identical, we expect the lubricant to act as an athermal solvent for the brush after the brush is saturated.}

					
					\section{Diffusion model derivation} \label{sec:diffusionModelDerivation}
					
					In the following, we assume that our system has azimuthal symmetry; in addition, we assume that the vertical advection in the brush is much faster than the lateral diffusion so that the concentration profile is homogeneous along the z-axis. The local concentration profile then only depends on the radial position $\rho$.
					
					\subsection{Free energy functional}
					
					To derive our dynamic equations, we need an expression for a free energy that
					depends on the local fraction of oil. We assume a functional
					form as the integral of a free energy density $f_B(\phi(x))$
					
					\begin{equation}
						\frac{\mathcal{F}_B[\phi(x)]}{k_B T} = \int \ud V \, f_B(\phi(x)).
					\end{equation}
					
					The free energy density of the brush can be written as
					
					\begin{equation}
						\label{eq:freeEnergyDensity_brush_SI}
						f_B(\phi(x)) = \frac{\phi}{N_o} \ln{\phi} + \frac{\chi(\phi)}{N_o} \phi \lrp{1-\phi} + \frac{k}{2} \frac{\sigma^2}{1-\phi}.
					\end{equation}
					
					One can easily verify that integrating this expression over the volume of the brush
					with a uniform oil fraction $\phi$ will result in the correct brush
					free energy Eq. \ref{eq:brushFreeEnergy_OG} but in the canonical
					ensemble instead of grand canonical. To do this we use Eq.
					\ref{eq:brushHeight_def} and (with $a=1$)
					
					\begin{align}
						&\phi V = n_o N_o \\
						&\int \ud V \frac{k}{2} \frac{\sigma^2}{1-\phi} = A \times H_B \frac{k}{2} \frac{\sigma^2}{1-\phi}\\
						&\sigma A = n_B
					\end{align}
					
					where $V$ and $A$ are the volume and area of the brush respectively. At equilibrium and for a given oil fraction $\phi^{\rm{eq}}$, the chemical potential of the brush is 
					
					\begin{equation}
						\label{eq:chemicalPotential_brush}
						\mu_B^{\rm{eq}} = \partDer{\mathcal{F}_B}{n_o} = \ln{\phi^{\rm{eq}}} + 1 - \phi^{\rm{eq}} + (1-\phi^{\rm{eq}})^2 \lrp{\chi + \chi' \phi^{\rm{eq}}} + k \frac{\sigma^2 N_o}{1-\phi^{\rm{eq}}}
					\end{equation}
					
					where we used
					
					\begin{equation}
						\partDer{\phi}{n_o} = \frac{\phi}{n_o} (1-\phi).
					\end{equation}
					
					In addition to the free energy of the brush, we account for the contribution of the three phase contact line through a free energy line-density $f_\rm{cl}(\lambda)$ where $\lambda$ is the line density of oil at the three phase contact line. With this, the free energy of the contact line $\mathcal{F}_\rm{cl}$ is given by
					
					\begin{equation}
						\frac{\mathcal{F}_\rm{cl} [\lambda]}{k_B T} = \oint \ud l \, f_\rm{cl}(\lambda)
					\end{equation}
					
					where the integral is evaluated around the contact line. We assume the free energy line-density to be quadratic in $\lambda$ and with a stiffness $\kappa$
					
					\begin{equation}
						\label{eq:freeEnergyDensity_contactLine_SI}
						f_\rm{cl}(\lambda) = \frac{\kappa}{2} \lrp{\lambda - \lambda_0}^2
					\end{equation}
					
					where $\lambda_0$ is the saturation line-density of the contact line. The chemical potential for the contact line is
					
					\begin{equation}
						\label{eq:chemicalPotential_contactLine}
						\mu_\rm{cl} = \partDer{f_\rm{cl}}{\lambda} = \kappa (\lambda - \lambda_0).
					\end{equation}
					
					The total free energy of the system can the be written as
					
					\begin{equation}
						\label{eq:freeEnergyFunctional_Total}
						\frac{\mathcal{F} [\phi(x), \lambda]}{k_B T} = \int \ud V \, \lrc{\frac{\phi}{N_o} \ln{\phi} + \frac{\chi(\phi)}{N_o} \phi \lrp{1-\phi} + \frac{k}{2} \frac{\sigma^2}{1-\phi}} + \oint \ud l \, \kappa \lrp{\lambda - \lambda_0}^2
					\end{equation}

					\subsection{Dynamical Equations}

					\subsubsection{Evolution of the line density}
					
					At the contact line, we assume that the rate of change of the line density of oil $\lambda$ is proportional to the difference in chemical potential and to the amount of lubricant in the brush at the location of the three phase contact line.
					
					\begin{equation}
						\label{eq:evolutionEq_lambda1_SI}
						\partDer{\lambda}{t} = -\mathcal{B} \Phi(R_{\rm{cl}}) \lrb{f'(\lambda) - \mu_B(R_{\rm{cl}},t)}
					\end{equation}
					
					where $\mathcal{B}$ is a filling rate, $\Phi(\rho) = \int_0^{H_B} \phi (\rho,z) \, \ud z$, $R_{\rm{cl}}$ is the radial position of the contact line, and $\mu_B(\rho,t)$ is a locally defined chemical potential for the brush. To simplify the picture, we assume that the chemical potential in the brush is always close to its equilibrium value from Eq. \ref{eq:chemicalPotential_brush}. Therefore, we rewrite the evolution equation Eq. \ref{eq:evolutionEq_lambda1_SI} for $\lambda$ as
					
					\begin{align}
						\label{eq:evolutionEq_lambdaFinal_SI}
						&\partDer{\lambda}{t} = -\mathcal{B} \Phi(R_{\rm{cl}}) \kappa \lrp{\lambda - \lambda_e}\\ \label{eq:lambda_equil_SI}
						&\lambda_e = \lambda_0 + \frac{\mu_B^{\rm{eq}}}{\kappa}
					\end{align}
					
					where we define the equilibrium line density $\lambda_e(\phi^{\rm{eq}})$ for a given equilibrium fraction of oil in the brush
					
					
					\subsubsection{Evolution of the concentration profile}
					
					We split our full domain into two regions separated by the contact line at $\rho=R_{\rm{cl}}$. In each region we define a current $j_\rho^{<}$ for the region $\rho < R_{\rm{cl}}$ and $j_\rho^{>}$ for $\rho > R_{\rm{cl}}$. At any position $\rho$, the current in the radial direction can be written as
					
					\begin{equation}
						\label{eq:diffusionCurrent_def_SI}
						j_\rho = -M \phi \nabla_\rho \frac{\delta \mathcal{F}_B \lrb{\phi}}{\delta \phi} 
					\end{equation}
					
					where $M$ is the mobility of oil in the brush, \rev{$\nabla _\rho = \partial/\partial \rho$} is the radial component of the gradient operator in cylindrical coordinates, and $\delta / \delta \phi$ denotes the variational derivative with respect to the concentration profile $\phi(\rho)$. From the current, the evolution equation for the profile can be obtained as the continuity equation
					
					\begin{equation}
						\partDer{\phi}{t} = - \nabla_\rho j_\rho.
					\end{equation}
					
					Using Eq. \ref{eq:freeEnergyDensity_brush} we can evaluate the variational derivative of our free energy functional
					
					\begin{align}
						\nonumber
						\frac{\delta \mathcal{F}_B \lrb{\phi}}{\delta \phi} &= \partDer{f_B}{\phi} \\ \label{eq:variationalDerivative_brush}
						&= \frac{\ln{\phi}}{N_o} + \frac{1}{N_o} + \frac{\chi'}{N_o} \phi \lrp{1-\phi} + \frac{\chi}{N_o} \lrp{1 - 2\phi} + \frac{k}{2} \frac{\sigma^2}{\lrp{1-\phi}^2}
					\end{align}
					
					and subsequently calculate the gradient
					
					\begin{align}
						\label{eq:variationalDerivative_gradient}
						\nabla_\rho \frac{\delta \mathcal{F}_B \lrb{\phi}}{\delta \phi} = \lrc{\frac{1}{N_o \phi} + k \frac{\sigma^2}{\lrp{1-\phi}^3} + \frac{1}{N_o} \lrb{\chi'' \phi\lrp{1-\phi} + 2\chi'\lrp{1-2\phi} - 2\chi}} \nabla_\rho \phi
					\end{align}
					
					which eventually gives the expression for the radial current
					
					\begin{align}
						\label{eq:diffusionCurrent_brush}
						j_\rho &= -M \phi \lrc{\frac{1}{N_o \phi} + k \frac{\sigma^2}{\lrp{1-\phi}^3} + \frac{1}{N_o} \lrb{\chi'' \phi\lrp{1-\phi} + 2\chi'\lrp{1-2\phi} - 2\chi}} \nabla_\rho \phi \\ \label{eq:diffusionCurrent_brush2}
						& = -M \lrc{\frac{1}{N_o} + k \frac{\sigma^2 \phi}{\lrp{1-\phi}^3} + \frac{\phi}{N_o} \lrb{\chi'' \phi\lrp{1-\phi} + 2\chi'\lrp{1-2\phi} - 2\chi}} \nabla_\rho \phi 
					\end{align}
					
					The final evolution equation for the concentration profile is therefore
					
					\begin{align}
						\nonumber
						&\partDer{\phi}{t} = M \Bigg\{ \lrb{\frac{1}{N_o} + k \frac{\sigma^2 \phi}{\lrp{1-\phi}^3} + \frac{\phi}{N_o} \lrb{\chi'' \phi\lrp{1-\phi} + 2\chi'\lrp{1-2\phi} - 2\chi}} \nabla_\rho^2 \phi \\ \label{eq:diffusionEq_brush}
						&+ \bigg[ k \sigma^2\frac{1 + 2\phi}{\lrp{1-\phi}^4} + \frac{1}{N_o} \lrp{\chi^{(3)} \phi^2\lrp{1-\phi} + \chi'' \phi \lrp{4-7\phi} + \chi' \lrp{2-10\phi} - 2\chi } \bigg] \lrp{\nabla_\rho \phi}^2 \Bigg\}
					\end{align}
					\
					
					To simplify the notation, we rewrite the current and the evolution in
					more compact forms. For the current we have
					
					\begin{align}
						\label{eq:diffusionCurrent_brush_compact}
						& j_\rho(\phi(t)) = -M \lrb{j_\rho^{0}(\phi(t)) + j_\rho^{\rm{int}}(\phi(t))} \nabla_\rho \phi \\ \label{eq:diffusionCurrent_brush_compact_0}
						& j_\rho^{0}(\phi(t)) = \frac{1}{N_o} + k \frac{\sigma^2 \phi}{\lrp{1-\phi}^3} \\ \label{eq:diffusionCurrent_brush_compact_int}
						& j_\rho^{\rm{int}}(\phi(t)) = \frac{\phi}{N_o} \lrb{\chi'' \phi\lrp{1-\phi} + 2\chi'\lrp{1-2\phi} - 2\chi}
					\end{align}
					
					and for the evolution equation we have
					
					\begin{align}
						\label{eq:diffusionEq_brush_compact}
						& \partDer{\phi}{t} = M \lrb{ g_1(\phi(t)) \lrp{\nabla_\rho \phi}^2 + g_2(\phi(t)) \nabla_\rho^2 \phi }\\ \label{eq:diffusionEq_brush_compact_g1}
						& g_1(\phi(t)) = g_1^{0}(\phi(t)) + g_1^{\rm{int}}(\phi(t)) \\ \label{eq:diffusionEq_brush_compact_g1_0}
						& g_1^{0}(\phi(t)) = k \sigma^2\frac{1 + 2\phi}{\lrp{1-\phi}^4} \\ \label{eq:diffusionEq_brush_compact_g1_int}
						& g_1^{\rm{int}}(\phi(t)) = \frac{1}{N_o} \lrp{\chi^{(3)} \phi^2\lrp{1-\phi} + \chi'' \phi \lrp{4-7\phi} + \chi' \lrp{2-10\phi} - 2\chi }\\ \label{eq:diffusionEq_brush_compact_g2}
						& g_2(\phi(t)) = g_2^{0}(\phi(t)) + g_2^{\rm{int}}(\phi(t)) \\ \label{eq:diffusionEq_brush_compact_g2_0}
						& g_2^{0}(\phi(t)) = j_\rho^{0}(\phi(t)) = \frac{1}{N_o} + k \frac{\sigma^2 \phi}{\lrp{1-\phi}^3} \\ \label{eq:diffusionEq_brush_compact_g2_int}
						& g_2^{\rm{int}}(\phi(t)) = j_\rho^{\rm{int}}(\phi(t)) = \frac{\phi}{N_o} \lrb{\chi'' \phi\lrp{1-\phi} + 2\chi'\lrp{1-2\phi} - 2\chi}
					\end{align}

					\subsubsection{Boundary conditions}
					
					To solve the evolution equations we need to set the appropriate boundary conditions at $\rho=0$, $\rho=R_{\rm{cl}}$, and $\rho=L$ with $L$ the length of the domain.\\
					
					At $\rho=0$, the azimuthal symmetry of our system implies
					
					\begin{equation}
						\label{eq:boundaryCondition_origin_SI}
						j_\rho^{<} (0,t) = 0
					\end{equation}
					
					since there is no preferred flow direction at the axis of symmetry.
					
					At $\rho=L$, we consider the brush to be in contact with an infinitely larger reservoir of oil that maintains the fraction at the edge to a preset value $\phi_B$
					
					\begin{equation}
						\label{eq:boundaryCondition_edge_SI}
						\phi(L,t) = \phi_B.
					\end{equation}
					
					Finally, at the contact line $\rho=R_{\rm{cl}}$, we require that all of the oil leaving the brush will be incorporated into the ridge/cloak. This means that the rate of change of $\lambda$ must match the total current coming from both sides of the contact line, which we write as
					
					\begin{equation}
						\label{eq:boundaryCondition_contactLine_SI}
						\partDer{\lambda}{t} = - H(\phi(R_{\rm{cl}})) \lrp{j_\rho^{>}(R_{\rm{cl}}) - j_\rho^{<}(R_{\rm{cl}})}
					\end{equation}


					\subsection{Numerical implementation}

					In order to solve the equation numerically, we split the domain radially into bins with width $\Delta r$, and label the separate bins as $r_i$ with $r_0=R$. For any function $g(r)$ we call $g_i \equiv g(r_i) $. We use the discretization scheme :
					
					\begin{align}
						&\frac{\partial g_i}{\partial r} = \frac{1}{2 \Delta r} ( g_{i+1} - g_{i-1} )\\
						&\frac{\partial^2 g_i}{\partial r^2} = \frac{1}{\Delta r^2} ( g_{i+1} + g_{i-1} - 2 g_{i} )
					\end{align}
					
					\noindent and the forward steps in time are executed as a forward Euler scheme.\\
					
					The boundary condition at the origin is imposed by setting
					
					\begin{equation}
						\label{eq:boundaryCondition_origin_num_SI}
						\phi_0(t) = \phi_1(t)
					\end{equation}
					
					while the boundary condition at the edge of the domain is imposed by maintaining for $\phi(L,t) \equiv \phi_L(t)$
					
					\begin{equation}
						\label{eq:boundaryCondition_edge_num_SI}
						\phi_L(t) = \phi_B
					\end{equation}
					
					The boundary condition at the contact line Eq. \ref{eq:boundaryCondition_contactLine_SI} is not as straightforward. The current on either side of the contact line is given by Eqs. \ref{eq:diffusionCurrent_brush_compact}-\ref{eq:diffusionCurrent_brush_compact_int}. However, to evaluate the derivative that appears there numerically, we use the backward and forward two-point formulas for the inner and outer domain respectively
					
					\begin{align}
						\label{eq:backwardDerivative}
						\nabla_\rho^{<} \phi |_{ \rho=R_{\rm{cl}} } = \frac{ \phi (R_{\rm{cl}}) - \phi(R_{\rm{cl}}-\Delta\rho) }{\Delta \rho} \\ \label{eq:forwardDerivative}
						\nabla_\rho^{>} \phi |_{ \rho=R_{\rm{cl}} } = \frac{ \phi (R_{\rm{cl}} + \Delta\rho) - \phi(R_{\rm{cl}}) }{\Delta \rho}.
					\end{align}
					
					Using Eq. \ref{eq:evolutionEq_lambdaFinal_SI}, \ref{eq:diffusionCurrent_brush_compact}, \ref{eq:backwardDerivative}, and \ref{eq:forwardDerivative}, and defining $\phi(R_{\rm{cl}},t)) \equiv \phi_{\rm{cl}}(t)$ we rewrite eq. \ref{eq:boundaryCondition_contactLine_SI} in the following way 
					
					\begin{align}
						\label{eq:boundaryCondition_contactLine_num1}
						-\mathcal{B} \Phi\lrp{R_{\rm{cl}}} \kappa (\lambda-\lambda_e) = H(R) M \lrc{ \lrb{j_\rho^{0}(\phi_{\rm{cl}}(t)) + j_\rho^{\rm{int}}(\phi_{\rm{cl}}(t))} } \times \\ \nonumber
						\frac{ \phi (R_{\rm{cl}} - \Delta\rho) - 2\phi_{\rm{cl}} + \phi (R_{\rm{cl}} + \Delta\rho)}{\Delta \rho}.
					\end{align}
					
					Given the value of $\phi$ at either side of the contact line, we can solve this equation for $\phi_{\rm{cl}}$. In particular, we assume that our brush is thin enough for the concentration profile to be independent of $z$, so that $\Phi(\rho) = H(\rho)\phi(\rho)$, and our boundary condition equation becomes
					
					\begin{align}
						\label{eq:boundaryCondition_contactLine_num}
						-\mathcal{B} \phi_{\rm{cl}} \kappa (\lambda-\lambda_e) = M  \lrb{j_\rho^{0}(\phi_{\rm{cl}}(t)) + j_\rho^{\rm{int}}(\phi_{\rm{cl}}(t))}  \times \\ \nonumber
						\frac{ \phi (R_{\rm{cl}} - \Delta\rho) - 2\phi_{\rm{cl}} + \phi (R_{\rm{cl}} + \Delta\rho)}{\Delta \rho}.
					\end{align}

					
					\section{Methods} \label{sec:methodsSI}

					
					\subsection{System preparation} \label{sec:systemPrep}
					
					\subsubsection{Liquid slab preparation}
					\label{sec:liquidPrep}

					Our systems of interest require the preparation of a liquid coexisting
					with a vapor phase in a slab geometry. The slab can consist of either
					a simple liquid or a polymeric liquid. We refer to the latter as a
					melt. The slab is prepared in two stages.  In the first stage, chains
					or individual particles are equilibrated in a box with periodic
					boundary conditions. The size of the box in the $x-y$ direction is
					chosen manually. The size in the z-direction is chosen so that the
					number density of monomers has a specific value. If the density of the
					liquid phase at coexistence is known, the starting density is chosen
					close to that. From experience, it is better to choose the starting
					density as slightly smaller than the coexistence value.  In the second
					stage the size of the simulation box in the z-direction is doubled.
					When simulating a polymeric liquid, chains that cross the boundary in
					the original box are properly unwrapped. The system is then left to
					equilibrate until a single liquid slab is coexisting with a dilute
					vapor phase.
					
					
					\subsubsection{Dry brush preparation}
					\label{sec:dryPrep}
					
					The polymer brush consists of chains of length $N_B$ grafted by one
					end by fixing the end monomer to one position. The position of the
					grafted monomer coincides with the purely repulsive flat surface. The
					brush has $n_B = N_x \times N_y$ chains, with $N_x$ and $N_y$ the
					number of grafting sites in the $x$ and $y$ directions respectively.
					The grafting sites are distributed on a square lattice with a lattice
					constant (distance between nearest sites) equal to $d$; this results
					in a grafting density
					
					\begin{equation}
						\sigma = \frac{1}{d^2}.
					\end{equation}
					
					The polymer brush is initialized with the chains fully elongated along
					the z-direction, then left to relax until equilibrium is reached. With
					the choice of parameters described in section
					\ref{sec:simulationModel}, the brush is in bad solvent and the
					equilibrium corresponds to a collapsed state with a rough topography.
					
					
					\subsubsection{Swollen brush preparation}
					\label{sec:swollenPrep}
					
					To swell the brush, it is placed in contact with a melt of oil chains
					prepared as described in section \ref{sec:liquidPrep}. The number of
					chains in the melt is $n_o$ and each chain has length $N_o$. The size
					of the melt in the $x-y$ direction is equal to that of the brush,
					which results in faster equilibration as the diffusion of free chains
					is quite slow. In the next step, we place the melt in the vicinity of
					the brush, and run the simulation until the brush absorbs the oil and
					swells to its equilibrium thickness. The number fraction $\Tilde{\Phi}$ of oil
					monomers to the total number of monomers is calculated as
					
					\begin{equation}
						\label{eq:oilNumFraction}
						\Tilde{\Phi} = \frac{ \rm{number~of~oil~monomers} }{ \rm{total~number~of~monomers} } = \frac{n_o N_o}{n_o N_o + n_B N_B}.
					\end{equation}

					
					\subsubsection{Liquid drop preparation and deposition} \label{sec:dropPrep}

					To prepare a drop to be deposited on the brush, the first step is to
					prepare a slab of simple liquid as described in section
					\ref{sec:liquidPrep}. The second step is to use this slab to deposit a
					droplet on the brush. To have good control over the contact radius and
					size of the droplet, we extract a hemisphere of radius $R$ from the
					slab and place it in contact with the brush, which is the starting
					point of our simulations.

					
					\subsection{Density maps} \label{sec:densityMapsCalc}
					
					For a particular simulation snapshot, the first step is to shift the
					coordinates of all particles so that the center of mass of the droplet
					coincides with the z-axis, while accounting for periodic boundary
					conditions. The density map $\varrho$ is created by counting the
					number of particles belonging to each bin and dividing by the volume
					of the bin. To avoid confusion, we will stick to the symbol
					$\varrho_{x}$ to denote density maps and $\rho_{x}$ to denote
					equilibrium bulk densities of species $x$. We will refer to the radial
					coordinate in cylindrical coordinates as $\rho$ with no subscript, and
					in spherical coordinates as $r$.
					
					\begin{figure}
						\centering
						\includegraphics[width=0.98\linewidth]{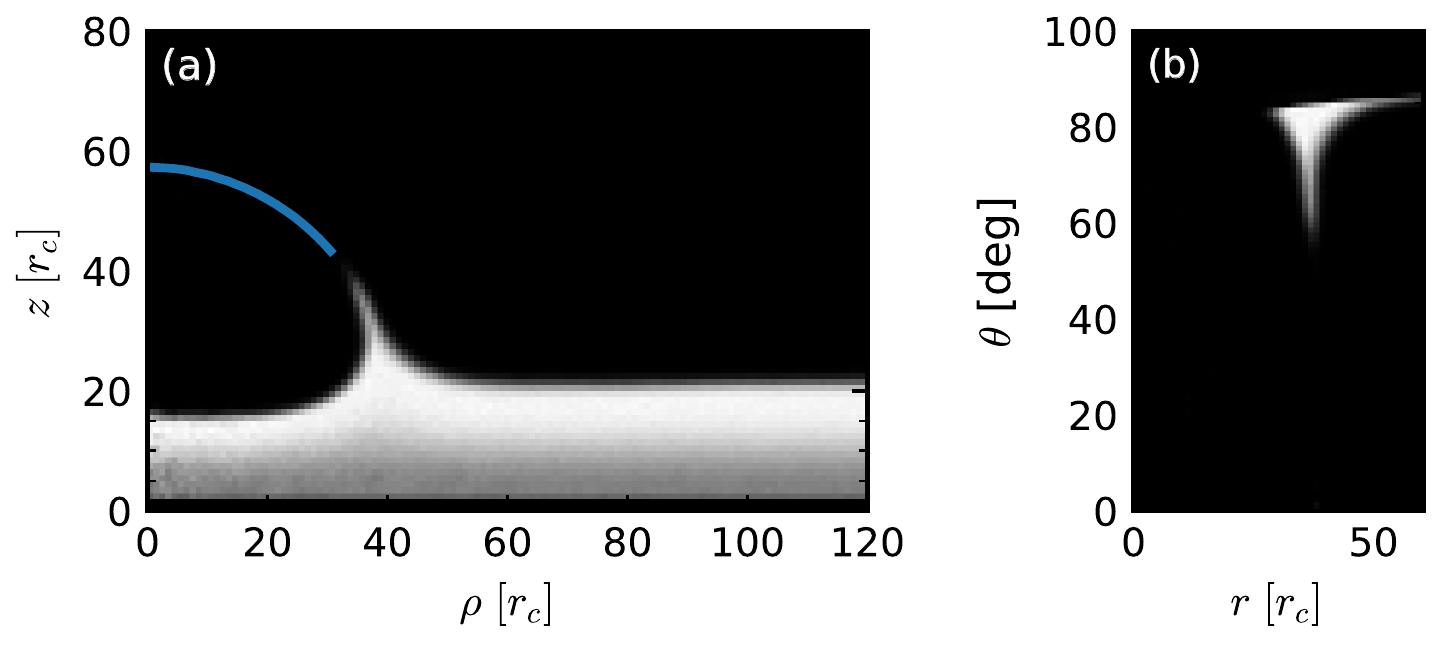}
						\caption{(a) Density map of oil in cylindrical coordinates $\varrho_{o}(\rho,z)$. Blue line indicates the droplet-vapor interface. (b) Density map of oil in spherical coordinates centered at the center of curvature of the drop $\varrho_{o}(r,\theta)$.}
						\label{fig:densityMaps_SI}
					\end{figure}
					
					
					\subsubsection{Cylindrical coordinates}
					
					The choice of cylindrical coordinates is motivated by the azimuthal
					symmetry of the drop. Therefore, the box is partitioned into bins in
					the z-direction and the radial direction, with respective sizes
					$\Delta z$ and $\Delta \rho$. The bin edges are positioned at
					
					\begin{align}
						\rho_i = i\times \Delta \rho \\
						z_j = j\times \Delta z
					\end{align}
					
					with $i,j \in \mathbb{N}$ the indices of the radial bin and of axial
					bins respectively. In this case, bins have different volumes depending
					on the index of the radial bin
					
					\begin{equation}
						\Delta V_{ij} = \pi \Delta z \Delta \rho^2 (2i+1).
					\end{equation}
					
					An example of such a density map $\varrho_o(\rho,z)$ for the oil is
					shown in Fig. \ref{fig:densityMaps_SI} (a).
					
					
					\subsubsection{Spherical coordinates}
					
					The origin of the coordinate system is placed at the center of
					curvature of the droplet. The box is partitioned into bins in the
					radial and polar directions, with respective sizes $\Delta r$ and
					$\Delta \theta$. The bin edges are positioned at
					
					\begin{align}
						r_i = i\times \Delta r \\
						\theta_j = j\times \Delta \theta
					\end{align}
					
					with $i,j \in \mathbb{N}$ the indices of the radial bin and of polar
					bins respectively. In this case, bins have different volumes depending
					on the index of the radial bin
					
					\begin{equation}
						\Delta V_{ij} = \frac{2 \pi}{3}\Delta r^3  (\cos{\theta_j} - \cos{\theta_{j+1}} ) [(i+1)^3 - i^3].
					\end{equation}
					
					An example of such a density map $\varrho_o(r,\theta)$ for the oil is
					shown in Fig. \ref{fig:densityMaps_SI} (b).
					
					
					\subsection{Equal density contours} \label{sec:densityContour}
					
					Having computed the density maps, one key method for delineating the
					boundaries between the different phases is through equal density
					contour lines.  An example of such a contour is shown in Figure
					\ref{fig:densityMaps_SI} (a), where the line is an equal density
					contour $\varrho_l(\rho, \phi) = \rho_l/2$, with $\rho_l$ the bulk
					density of the bulk polymer phase. Density contours are used in
					various ways to extract information from the simulations as described
					below.
					
					\subsubsection{Droplet shape}
					
					One way of using the density contours is to characterize the shape of
					the droplet. In the case of a sessile droplet, a density contour at
					$\varrho_l(\rho,z) = \rho_l/2$ is extracted for the density map of the
					droplet; afterwards, the top half is fit to a circle, providing us
					with the coordinates of the center of curvature and the radius of
					curvature of the droplet. This information is then used to calculate
					the spherical density maps $\varrho_x(r,\theta)$.
					
					
					\subsubsection{Unperturbed brush height}
					
					To find the height of the brush far away from the droplet, we
					calculate density contour $\varrho_o(\rho,z) = \rho_o/2$ from the
					density map of oil, where $\rho_o$ is the density of the bulk polymer
					phase. From this contour, we then calculate the average height far
					away from the drop, and use that as the height of the unperturbed
					brush. This height is used as the reference to calculate the height of
					the wetting ridge and the apparent contact angles.
					
					
					\subsubsection{Apparent contact angle} 
					
					Knowing the height of the brush away from the droplet, we can use it as
					a baseline for calculating apparent contact angles. In our work, the
					apparent contact angles are always defined as the angle the fit to the
					droplet shape makes with a horizontal line at the height of the
					unperturbed brush
					
					\begin{equation}
						\theta_{\rm{app}} = \frac{\pi}{2} - \arcsin\lrp{\frac{z_B - z_c}{R_{\rm{D}}}}
					\end{equation}
					
					where $z_B$ is the z-coordinate of the surface of the unperturbed
					brush, $z_c$ is the z-coordinate of the center of curvature of the
					droplet, and $R_{\rm{D}}$ is the radius of curvature of the droplet.
					
					
					\subsubsection{Ridge height} 
					
					To calculate the height of the ridge, we calculate the density contour
					$\varrho_o(\rho,z) = \rho_o/2$ in the presence of a droplet. The apex
					of the ridge is selected as the point with the largest z-coordinate
					$z_{\rm{max}}$. The height of the ridge is then calculated as 
					
					\begin{equation}
						h(t) = z_{\rm{max}}(t) - z_B(t)
					\end{equation}
					
					where $z_B$ is the z-coordinate of the oil in the brush far away from the droplet.
					

					
					\subsection{Calculating the thickness of the cloak} \label{sec:cloakThicknessCalc}
					
					To quantify the thickness of the cloak with time, we use the relation
					
					\begin{equation}
						\label{eq:cloakThickDer1}
						N = \int \varrho_{o} r^2 \sin\theta \, \ud r \, \ud \theta \, \ud \phi
					\end{equation}
					
					where $N$ is the number of oil monomers in the integration volume and
					$\varrho_{o}(r,\theta,\phi)$ is the local density of oil in spherical
					coordinates centered at the center of curvature of the droplet. Since
					the cloak thickness is much smaller than the drop radius of curvature
					we set $r=R_{\rm{D}}$, and since we have azimuthal symmetry, we
					rewrite Eq. \ref{eq:cloakThickDer1} as
					
					\begin{align}
						\nonumber
						N(t) &= 2\pi R_{\rm{D}}^2 \int_{\theta_{\text{tip}}(t)}^{\theta_{\text{ridge}}} \int_{0}^{\infty} \varrho_{o} \sin\theta \, \ud r \, \ud \theta \\ \label{eq:cloakThickDer2} 
						&= 2\pi R_{\rm{D}}^2 \int_{\theta_{\text{tip}}(t)}^{\theta_{\text{ridge}}} \Sigma \sin\theta \, \ud \theta.
					\end{align}
					
					where $\theta_{\text{tip}}$ is the angular position of the cloak front
					and $\theta_{\text{ridge}}$ is the angular position of the wetting
					ridge, and we define $\Sigma=\int_0^{\infty} \varrho_{o} \, \ud r$.
					The ridge remains more or less unchanged as the cloaking progresses;
					therefore, we determine $\theta_{\text{ridge}}$ manually at
					$\theta_{\text{ridge}} = 65^{\circ}$ from the $\Sigma(\theta)$ curve
					at the highest swelling ratio, and use this value for all lower
					swelling ratios. Given the number of oil monomers, the thickness of
					the cloak can be written as \cite{platikanov2005thin}
					
					\begin{equation}
						\label{eq:cloakThickDer3}
						h = \frac{N}{\rho_o A}
					\end{equation}
					
					with $\rho_o$ and $A$ the bulk density of oil and the area covered by
					the cloak respectively. The area can be calculated through
					
					\begin{equation}
						\label{eq:cloakThickDer4}
						A(t) = 2\pi R_{\rm{D}}^2 ( \cos\theta_{\text{tip}}(t) - \cos\theta_{\text{ridge}} )
					\end{equation}
					
					Finally, using Eqs. \eqref{eq:cloakThickDer2},
					\eqref{eq:cloakThickDer3}, and \eqref{eq:cloakThickDer4}, we can
					calculate the thickness of the cloak from $\Sigma(\theta,t)$ as
					
					\begin{equation}
						h(t) = \frac{\int_{\theta_{\text{ridge}}}^{\theta_{\text{tip}}(t)} \Sigma \sin\theta \, \ud\theta}{\rho_o \left[ \cos\theta_{tip}(t) - \cos\theta_{ridge} \right]}.
					\end{equation}
					
					
					\subsection{Calculating the interaction function from simulations}\label{sec:brushFreeEnergy_variableChiCalc}

\begin{figure}[t]
						\begin{center}
							\includegraphics[width=0.5\textwidth]{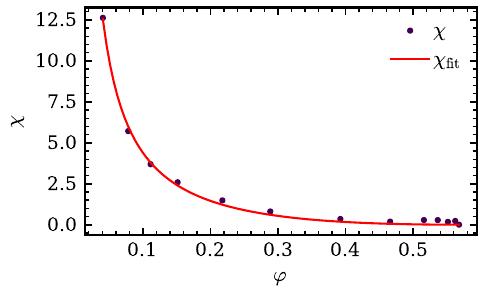}
						\end{center}
						\caption{The swelling dependent interaction function $\chi(\phi)$ as calculated from our simulation data. The red line is a fit to the function \ref{eq:fittingFunction}.}
						\label{fig:interactionParameterVphi}
					\end{figure}

					To calculate the interaction function from Eq.
					\ref{eq:variableInteractionFunction} we need to evaluate $
					\Tilde{\mathcal{F}}_0(\phi, \mu)$ which depends on the chemical
					potential in the brush. When our brush is at equilibrium, it coexists
					with a vapor phase of oil chains. This means that the chemical
					potential in the brush is equal to that of the vapor. Keeping in mind
					that our reference chemical potential is that of a saturated brush, we
					can find the chemical potential of an undersaturated brush through
					
					\begin{equation}
						\label{eq:chemicalPotential_vapor}
						\mu_B(\phi)=k_BT \ln{ \frac{ \rho_{\rm{v}}(\phi) }{ \rho_{\rm{v}}^{\rm{sat}} } }
					\end{equation}
					
					with $\rho_{\rm{v}} (\phi)$ the density of the vapor of polymers,
					$\rho_{\rm{v}}^{\rm{sat}}$ the vapor density for a saturated brush. We
					then use Eq. \ref{eq:variableInteractionFunction} to calculate
					$\chi(\phi)$. For the fraction of oil $\phi$ we use the definition
					
					\begin{equation}
						\phi = 1 - \frac{1}{\alpha}
					\end{equation}
					
					where $\alpha=H_B/H_0$ is the swelling ratio of the brush. The
					resulting $\chi(\phi)$ is shown in Figure
					\ref{fig:interactionParameterVphi}. To implement the interaction
					function in our diffusion model we choose to fit the simulation data
					to an ad-hoc function of the form
					
					\begin{equation}
						\label{eq:fittingFunction}
						\chi(\phi) = A \frac{\lrp{\phi^* - \phi}^2}{\phi^\beta}.
					\end{equation}
					
					After fitting, we find the values for the parameters
					
					\begin{equation}
						A=2.48 \pm 0.28 \, ; \, \beta = 0.90 \pm 0.04
					\end{equation}
					
					where the errors are obtained from the covariance matrix of the fit.
					
					While the fit is not perfect, it provides a good starting point to see
					what a variable $\chi$ model can achieve with the diffusion equation.
					In the diffusion model, some derivatives of the interaction function
					appear, which we write through our fit function as
					
					\begin{align}
						&\chi'(\phi) = -\frac{\beta}{\phi} \chi - 2 \frac{A}{\phi^{\beta}} \lrp{\phi^* - \phi}\\
						&\chi''(\phi) = -\frac{\beta}{\phi} \chi' + \frac{\beta}{\phi^2} \chi  + 2 \beta \frac{A}{\phi^{\beta+1}} \lrp{\phi^*-\phi} + 2 \frac{A}{\phi^\beta} \\
						&\chi^{(3)}(\phi) = -\frac{\beta}{\phi} \chi'' + 2\frac{\beta}{\phi^2} \chi' - 2\frac{\beta}{\phi^3} \chi - 2 \beta (\beta+1) \frac{A}{\phi^{\beta+2}}(\phi^*-\phi) - 4 \beta \frac{A}{\phi^{\beta+1}}
					\end{align}
					

					\subsection{Determining continuum model parameters} \label{sec:modelParametersCalc}
					
					In our dynamical equations above, we have many undetermined
					parameters, some of which we introduce heuristically such as the
					saturation line density $\lambda_0$ or the brush elastic constant $k$.
					Many of our model parameters can be determined from the results of our
					simulations, as described in this section.
					
					\subsubsection{Saturation fraction} 
					
					To determine the saturation fraction, we measure the height of the
					brush $H_B$ from the simulations, from that we calculate the fraction
					of oil as
					
					\begin{equation}
						\label{eq:modelParamDet_saturationPhi}
						\phi = 1 - \frac{H_0}{H_B}.
					\end{equation}
					
					When the brush is oversaturated with oil, the value calculated from
					the above equation will correspond to the saturation fraction
					$\phi^*$. We find for PDMS-like brushes $\phi^* \approx 0.57$ and for
					PLMA-like brushes we find $\phi^* \approx 0.59$.

					\begin{figure}[h]
							\begin{center}
								\includegraphics[width=7cm]{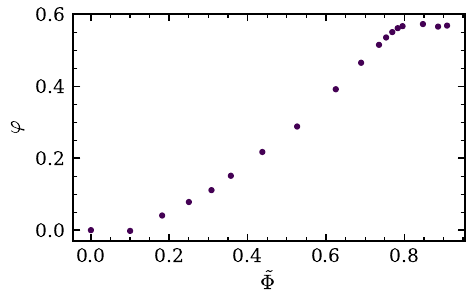}
								\includegraphics[width=7cm]{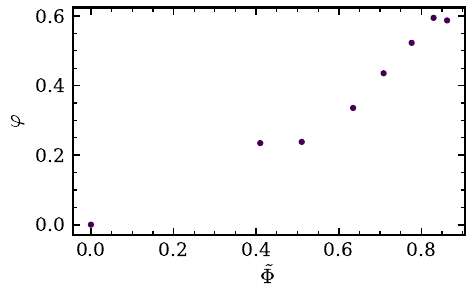}
							\end{center}
							\caption{The lubricant fraction $\phi$ calculated from Eq.
								\ref{eq:modelParamDet_saturationPhi} versus the number fraction of
								oil in the system $\Tilde{\Phi}$ as calculated from Eq.
								\ref{eq:oilNumFraction}. The values at large $\phi$ correspond to
								the saturation fraction. (a) is the result for PDMS-like brushes
								with $\phi^* \approx 0.57$, (b) for the PLMA-like brushes with
								$\phi^* \approx 0.59$.}
							\label{fig:realPhi}
						\end{figure}
						
						\subsubsection{Elastic constant} 
						
						According to Eq. \ref{eq:variableInteractionFunction}, the
						interaction part of the free energy vanishes when the brush is
						saturated at oil fraction $\phi^*$. In addition, the chemical
						potential of the brush vanishes at saturation since we use the
						saturated brush as a reference. Therefore, the elastic constant of the
						brush can be determined by setting
						
						\begin{align}
							\label{eq:modelParamDet_elastic1}
							\mu_B^{\rm{eq}} (\phi^*) = 0 = \ln{\phi^*} + 1 - \phi^* + k \frac{\sigma^2 N_o}{1-\phi^*}.
						\end{align}
						
						Since $\phi^*$ can be determined from the simulation, the only unknown is $k$. 
						
						\begin{equation}
							\label{eq:modelParamDet_elasticFinal}
							k = -\frac{1-\phi}{\sigma^2 N_o} \lrp{\ln{\phi^*} + 1 - \phi^*}
						\end{equation}
						
						We find for PDMS-like brushes $k=0.182$ and for PLMA-like brushes $k=0.154$
						
						\subsubsection{Height of dry brush} 
						
						The height of the dry brush can be easily extracted from the
						simulations with no oil. For PDMS-like brushes, one gets $H_0=9.431 \,
						[r_c]$, and for PLMA-like brushes, $H_0=9.872 \, [r_c]$.
						
						\subsubsection{Equilibrium line density} 
						
						The value of $\lambda_e$ can be determined from simulation by taking
						the total number of monomers in the ridge/cloak at equilibrium
						$N_{\rm{cl}}$, and dividing the result with the circumference of the
						contact line
						
						\begin{equation}
							\label{eq:modelParamDet_lambdaEquil}
							\lambda_e = \frac{N_{\rm{cl}}}{2 \pi R_{\rm{cl}}}.
						\end{equation}
						
						In practice, we average the last 3 points in out $\lambda(t)$ time-series.
						
						\subsubsection{Contact line free energy density} 
						
\begin{figure}[h]
								\begin{center}
									\includegraphics[width=7cm]{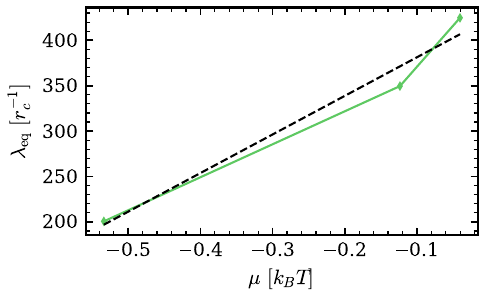}
									\includegraphics[width=7cm]{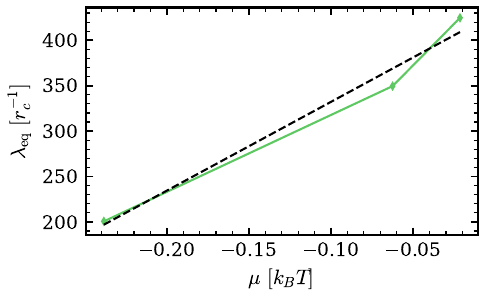}
								\end{center}
								\caption{Equilibrium values for the line density $\lambda_e$ versus the chemical potential in the brush calculated using (a) variable interaction function $\chi(\phi)$ and (b) vanishing interaction function $\chi=0$. The dashed lines are linear fits.}
								\label{fig:lambdaEqFit}
							\end{figure}

						The parameters $\lambda_0$ and $\kappa$ can be calculated by fitting
						Eq. \ref{eq:lambda_equil_SI} for different oil fractions. The value of
						$\lambda_e$ can be calculated as described in this section, while the
						chemical potential can be calculated using Eq.
						\ref{eq:chemicalPotential_brush}. The results are shown in Figure
						\ref{fig:lambdaEqFit} for variable interaction function in (a) and for
						$\chi=0$ in (b). We find for PDMS-like brushes $\lambda_0\approx423.8
						~ ; ~ \kappa \approx 2.35\times10^{-3}$ for variable interaction
						function and $\lambda_0\approx430 ~ ; ~ \kappa \approx
						1.02\times10^{-3}$ for $\chi=0$.

							\clearpage
							
							\section{Additional Data on Oil Separation} \label{sec:oilSeparationSI}
							
\begin{figure}[h]
								\begin{center}
									\includegraphics[width=7cm]{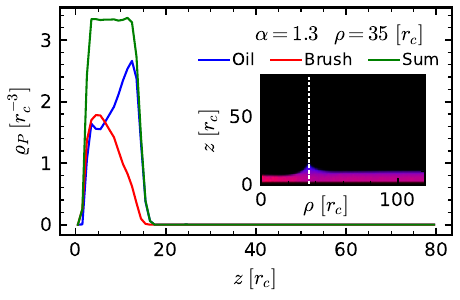}
									\includegraphics[width=7cm]{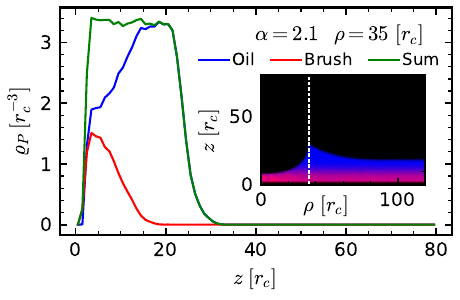}
									\includegraphics[width=7cm]{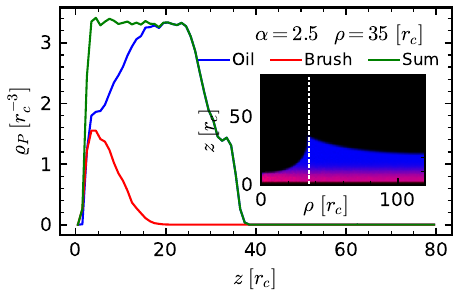}
								\end{center}
								\caption{Vertical density profile of polymers in the W-H system in
									the area of the wetting ridge at radial position $\rho=35 \,
									[r_c]$ for swelling ratios (a) $\alpha=1.3$, (b) $\alpha=2.1$ and
									(c) $\alpha=2.5$ (oversaturated). Insets show the density map
									$\varrho(\rho,z)$ with the grafted chains in red and the oil in
									blue. Magenta indicates a mixture of the two. The dashed lines
									indicate the position where the density profile is shown. For the
									higher swelling ratios we clearly see that the fluid separates
									from the brush as indicated by a density of oil equal to the bulk
									density of polymers.}
								\label{fig:fluidSeparation_PLMA_SI}
							\end{figure}

							In this section we present additional results for the separation of
							oil from the brush in the wetting ridges of the W-H system. Figure
							\ref{fig:fluidSeparation_PLMA_SI} shows the density profile of grafted
							chains, oil chains, and the sum of the two. The insets show the full
							density map of polymers $\varrho_P(\rho,z)$ with the grafted chains
							labeled in red and the oil in blue. The magenta color indicates
							regions where the two are mixed, and the dashed lines show our choice
							of the position of the wetting ridge at $\rho=35 \, [r_c]$. We can
							clearly see that for the low swelling ratio there is not separation
							since the density of oil is always lower than the sum. Close to
							the saturation point, at $\alpha=2.1$, the oil separates from the
							ridge.  For the oversaturated brush, separation is also observed
							as expected.
							
\begin{figure}[h]
								\begin{center}
									\includegraphics[width=7cm]{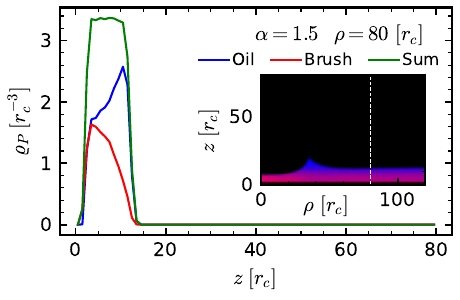}
									\includegraphics[width=7cm]{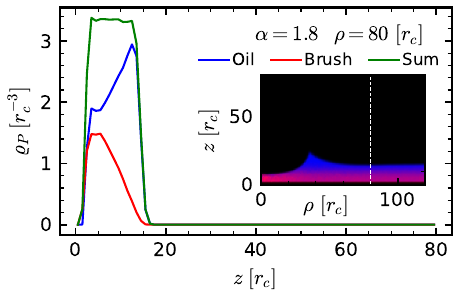}
									\includegraphics[width=7cm]{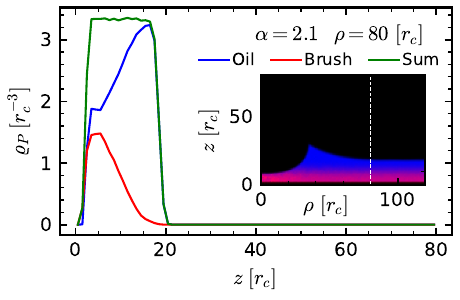}
								\end{center}
								\caption{Similar to Fig. \ref{fig:fluidSeparation_PLMA_SI} but at $\rho=80 \, [r_c]$ for swelling ratios (a) $\alpha=1.5$, (b) $\alpha=1.8$ and (c) $\alpha=2.1$.}
								\label{fig:fluidSeparation_PLMA_SI_far}
							\end{figure}

							To confirm that the fluid separation is due to the drop and localized
							in the wetting ridge, we also look at the density profiles of the
							grafted chains, oil chains, and the sum far away from the ridge at
							$\rho=80 \, [r_c]$. We only show the profiles for undersaturated
							brushes or close to saturation. The results are shown in Figure
							\ref{fig:fluidSeparation_PLMA_SI_far} where it is clear that no
							separation occurs far away from the ridge.
							
							\clearpage
							
							\section{Contacting drops} \label{sec:contactingDropsSI}
							
\begin{figure}[h]
								\begin{center}
									\includegraphics[width=0.7\textwidth]{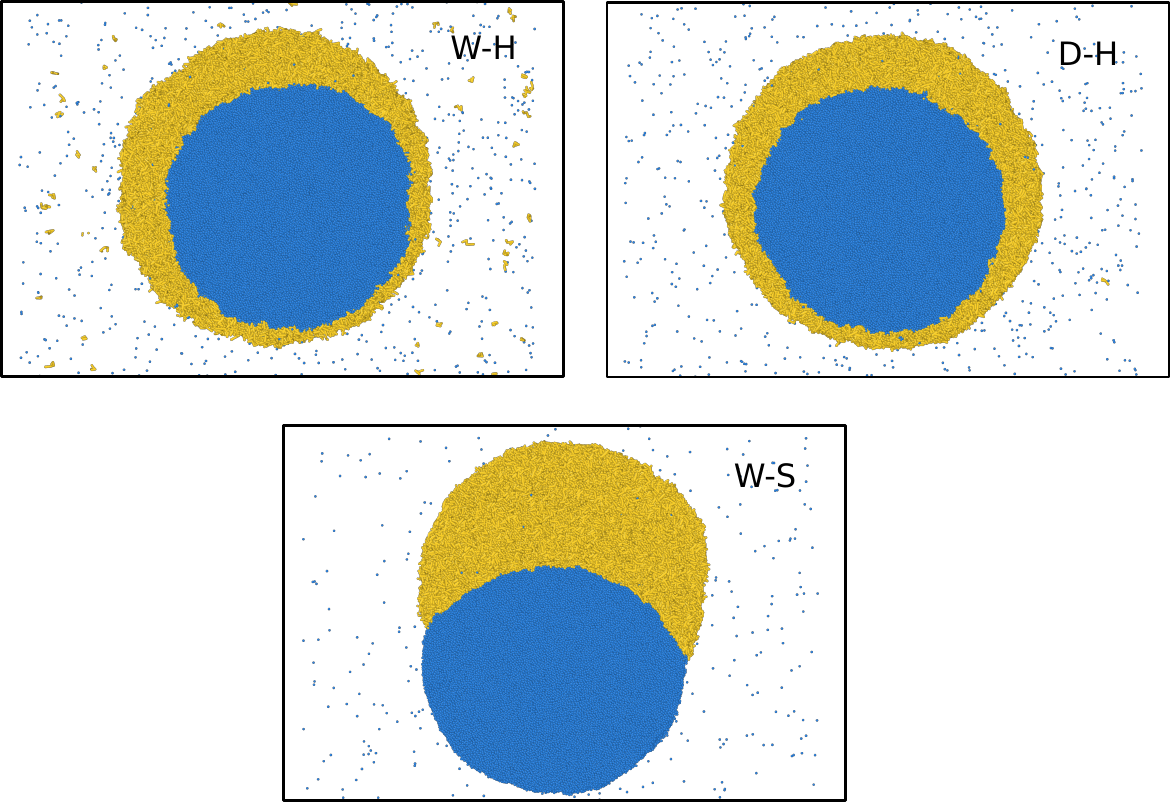}
								\end{center}
								\caption{Snapshots from simulations of the systems analogous to
									water on PLMA, DMSO40 on PLMA, and water on PDMS. Yellow chains
									are oil, red are grafted chains, and blue particles constitute the
									liquid. \rev{The simulation time is $6.4\times 10^3 [\tau]$.} }
								\label{fig:snapshot_contactingDrops_SI}
							\end{figure}

							The snapshots in Fig. \ref{fig:snapshot_contactingDrops_SI} show the
							final configurations after $6.4\times 10^3 \, [t]$ in simulations of
							contacting droplets of the liquid and the oil for the three systems we
							investigated. We clearly see that the oil cloaks the liquid in the W-S
							and D-H systems, where the spreading parameter of oil on the liquid is
							positive. In the W-H system the configuration is that of two fused
							immiscible droplets.
							
							\clearpage
							\section{Oil distribution at lower swelling} \label{sec:oilDistributionSI}
							
\begin{figure}[h]
									\begin{center}
										\includegraphics[width=7cm]{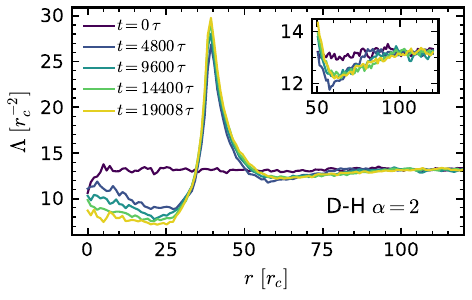}
										\includegraphics[width=7cm]{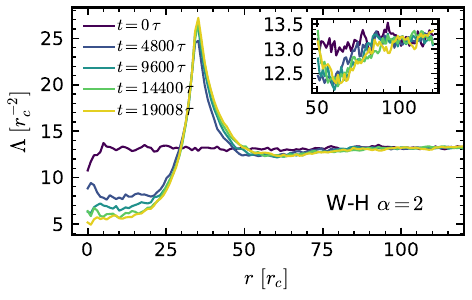}
									\end{center}
									\caption{$\Lambda(\rho)$ at different time points during cloaking
										for the (a) D-H, (b) W-H, and (c) W-S systems at $\alpha=2$.
										Insets are close ups near the three phase contact line. There is a
										clear depletion zone of material outside of the drop. The amount
										of material also drops from under the droplet.}
									\label{fig:LambdaVsTime_lowSwelling}
								\end{figure}

							Fig. \ref{fig:LambdaVsTime_lowSwelling} shows the oil distribution
							during ridge growth at low swelling ratio $\alpha=2$. 
							The depletion zone is localized to the vicinity of the wetting
							ridge, as opposed to the case of higher swelling ratios, where it
							extends further.
							
								\clearpage
								\section{Comparison of theoretical models} \label{sec:compareTheorySI}
								
\begin{figure}[h]
										\begin{center}
											\includegraphics[width=0.5\textwidth]{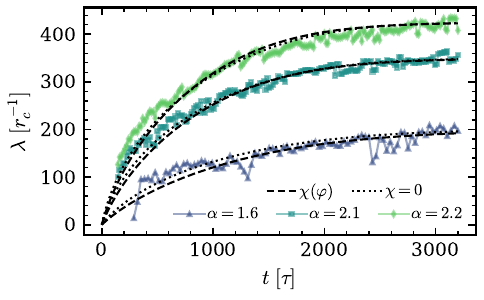}
										\end{center}
										\caption{The line density at the ridge from simulation (data
											points), theory with variable interaction term (dashed lines), and
											theory with no interaction term (dotted lines).}
										\label{fig:lineDensity_WS_compare}
									\end{figure}

								Fig. \ref{fig:lineDensity_WS_compare} shows the simulation data for
								the line density at the three phase contact line along with the
								results of the diffusion model with variable interaction function
								$\chi(\phi)$ or with vanishing interaction term $\chi=0$.
								
								\end{document}